\documentclass[onecolumn]{autart}    

\usepackage{savesym}
\savesymbol{AND}
\usepackage{graphicx,subfigure}          
\usepackage{amsmath} 
\usepackage{amssymb}  
\usepackage{amsfonts, bm}
\usepackage{algorithm,algorithmic}
\usepackage{arydshln,leftidx,mathtools,multirow}
\usepackage[dvips]{epsfig}   
\usepackage{psfrag,url}

\renewcommand{\thefigure}{\arabic{figure}}

\newcommand{\algrule}[1][.2pt]{\par\vskip.5\baselineskip\hrule height #1\par\vskip.5\baselineskip}
\newcommand{\algrrule}[1][.1pt]{\par\vskip.5\baselineskip\hrule height #1\par\vskip-.5\baselineskip}

\newtheorem{definition}{\hspace{-1em}{\textit{Definition}}}

\newtheorem{remark}{\hspace{-1em}{\textit{Remark}}}

  \usepackage[color]{changebar}
\cbcolor{red}
\usepackage{tikz}
\usepackage{verbatim}
\usetikzlibrary{trees,positioning}

\renewcommand{\thefigure}{\arabic{figure}}

\begin{document}

\begin{frontmatter}

\title{Efficient hinging hyperplanes neural network and its application in nonlinear system identification} 

\thanks[footnoteinfo]{This paper was not presented at any IFAC 
meeting. Corresponding author Xiangming Xi.}

\author[Paestum]{Jun Xu}\ead{xujunqgy@hit.edu.cn},    
\author[Rome,Baiae]{Qinghua Tao}\ead{taoqh14@mails.tsinghua.edu.cn},               
\author[Paestum]{Zhen Li}\ead{1654995440@qq.com},  
\author[Paestum]{Xiangming Xi} \ead{xixiangming@gmail.com},
\author[Baiae]{Johan A. K. Suykens} \ead{johan.suykens@esat.kuleuven.be},
\author[Rome]{Shuning Wang}\ead{swang@tsinghua.edu.cn}

\address[Paestum]{School of Mechanical Engineering and Automation, Harbin Institute of Technology, Shenzhen, 518055, China}  
\address[Rome]{BNRist, Department of Automation, Tsinghua
University, Beijing, 100084, China}             
\address[Baiae]{STADIUS, ESAT, KU Leuven, Leuven, 3001, Belgium}        


\begin{abstract}                          
In this paper, the efficient hinging hyperplanes (EHH) neural network is proposed, which is basically a single hidden layer neural network. Different from the dominant single hidden layer neural networks, the hidden layer in the EHH neural network can be seen as a directed acyclic graph (DAG) and all the nodes in the DAG contribute to the output.  It is proved that for every EHH neural network, there is an equivalent adaptive hinging hyperplanes (AHH) model, which was proposed  based on the model of hinging hyperplanes (HH) and finds good applications in system identification. Analog to the proof for the AHH model, the universal approximation ability of the EHH neural network is provided. Different from other neural networks, the EHH neural network has interpretability ability, which can be easily obtained through its ANOVA decomposition (or interaction matrix). The interpretability can then be used as an indication for the importance of the input variables. The construction of the EHH neural network includes initial network generalization and parameter optimization (including the structure and weights parameters). A descent algorithm for searching the locally optimal EHH neural network is proposed and the worst-case complexity of the algorithm is also provided. The EHH neural network is applied in nonlinear system identification, the simulation results show that satisfactory accuracy can be achieved with relatively low computational cost, and at the same time, some insights into the regressors and the interactions among the regressors can be revealed.
\end{abstract}

\end{frontmatter}

\section{Introduction}

Nonlinear system identification is an important procedure to predict future behavior of a dynamic system, to apply model-based control or to understand the physical insights. Basically,  the identification problem is to infer relationships between the system output and its past input-output data. Collect a finite number of past inputs $u(k)$ and outputs $y(k)$ into the regressor $\varphi(k)=[y(k-1), \ldots, y(k-n_b ), u(k), u(k-1), \ldots, u(k-n_a)]^T$, then the problem  is to understand the relationship between the next output $y(k)$ and $\varphi(k)$ \cite{Sjoberg1994}.   Since most of the practical systems in the real world are so complicated that the analytical models can not be established at low cost, the nonlinear black-box models are often used to depict the nonlinear system, of which nonlinear neural networks find wide applications \cite{Narendra1990,Suykens2012artificial}. Ever since its emergence, the neural network has gained a lot of popularity, especially when it was proved that with enough hidden nodes, a neural network model can approach almost all continuous systems  at arbitrary precision \cite{Hornik1989multilayer}. However, the neural network also receives criticism as it lacks interpretability, meaning that using it as a black-box model to depict a nonlinear system can not reveal the intrinsic mechanism.

The piecewise linear (PWL) neural network is a special kind of neural networks, which admits a linear relationship in each subregion. In PWL neural networks, the nonlinearity only comes from the PWL activation functions. For example, the fully connected neural network which employs the popular rectified linear units (ReLU), $\max\{0, x\}$, is a kind of PWL neural networks, since ReLU is exactly the basis function of the hinging hyperplanes (HH) model \cite{Breiman1993hinging}. A second example goes to the maxout neural networks \cite{Goodfellow2013maxout}, which lays its foundation on the generalized hinging hyperplanes (GHH) model proposed in \cite{Wang2005}. Both the HH model and GHH model belong to the family of continuous PWL representations and they differ in the sense of representability, i.e., the HH model can not represent all the continuous PWL functions in higher than 2 dimensions while the GHH model can. Besides, another member in the family is the  adaptive hinging hyperplanes (AHH) model, which was proposed to eliminate the inefficiency caused by the gradient-based method in the identification of the HH and GHH models \cite{Xu2009ahh}.  The AHH model is derived using a recursive partitioning procedure with a greedy strategy, i.e., by traversing all existing basis functions, the  basis function that yields the largest decrease in the optimized criterion is added to the model in the traversal manner. The AHH model has been successfully applied in regression, optimization and dynamic system identification \cite{Huang2010operation,Xu2012model,Gao2015optimizing}. However, the application of the AHH model to problems with a higher dimension is limited by two reasons. The first is the identification speed, which is exponential with respect to the number of basis function \cite{Xu2009ahh}, and the increase is more abrupt when the dimension is higher. The second owes to the redundancy and duplication in the calculation of  the same factors that appear more than once in the basis functions, thus the AHH model can not be  realized as a distributed representation, which fails to be accelerated by clusters of computers.

From the structural point of view, the HH, GHH and AHH models can be regarded as single hidden layer PWL neural networks, and the weighted sum of the nodes that represent the basis functions constitute the output of the neural network. Different from the existing neural networks that based on the HH and GHH models, we investigate a variant of the AHH structure that is essentially more extendable in applications. We name the proposed structure as the efficient hinging hyperplanes (EHH) neural networks for high-dimensional problems, which only employ the ``$\min$'' and ``$\max$" activation functions. It is basically an \textbf{interpretable} one hidden layer PWL neural network, in which the hidden layer is modelled as a directed acyclic graph (DAG). Each EHH neural network corresponds to an  AHH model. Moreover, the EHH representation has a distributed essence which the AHH representation fails to have. 

The remainder of this paper is organized as follows: Section 2 gives a review of PWL models based on the model of HH. Section 3 describes the structure, the approximation ability as well as the interpretability of the EHH neural network, besides, the relationship between the AHH model and the EHH neural network is provided. Then the training of the EHH neural network is performed in Section 4, in which both the structure and the weights of the neurons are optimized. Section 5 illustrates the application of the EHH neural network to nonlinear system identification. Finally the paper ends with conclusions and future work in Section 6.

\emph{Notations}. A vector $\mathbf{x}$ is written in bold, while a matrix is denoted by the capital bold, like $\mathbf{Z}$. The notation $x_i$ denotes the $i$-th component in $\mathbf{x}$, while the notation $\mathbf{Z}_{ij}$ denotes the element in the $i$-th row, $j$-th column of $\mathbf{Z}$, and $\mathbf{Z}(:, i)$ denotes the $i$-th column in the matrix $\mathbf{Z}$. The superscript ``$T$" denotes the transpose. 
The real scalar set is denoted by $\mathbb{R}$ while the real vector set in $n$ dimension is written as $\mathbb{R}^n$. The notations $\mathrm{var}$ and $\sigma$ denote the variance and standard deviation for a data vector, respectively. The italic capital, $A$, represents the nodes in the hidden layer, and $D$ and $C$ stand for the source nodes and intermediate nodes, respectively. The calligraphic capitals stands for sets.

\section{Review of the HH family}
The model of HH was first proposed by Breiman \cite{Breiman1993hinging}, and can be cast as a linear combination of basis functions, i.e.,
\begin{equation}\label{basis-expansion}
f_{\rm{HH}}=\sum\limits_{m=1}^M a_m B_m(\mathbf{x}),
\end{equation}
in which $\mathbf{x} \in \mathbb{R}^n$ is the input and $f_{\mathrm{HH}} \in \mathbb{R}$ is the output. In the HH model, the basis function $B_m(x)$ takes the form of
\begin{equation}\label{hinge}
\max\{0, \ell_m(\mathbf{x})\},
\end{equation}
 in which $\ell_m(\mathbf{x})$ is a linear (affine) function \cite{Breiman1993hinging} (Here we use ``linear'' to represent both linear and affine functions). The expression (\ref{hinge}) is called a hinge.  It is worthy to note that the commonly used ReLU activation function in deep networks is a special kind of the hinge by restricting the linear function $\ell_m(\mathbf{x})$ to be univariate affine.
 The model of HH has been applied quite successfully in regression, classification \cite{Huang2013support}, and dynamic system identification \cite{Chikkula1998}.
 
As the model of HH can not represent some continuous PWL functions in 2 and higher dimensions, the model of GHH was proposed to generalize the model of HH by replacing the hinge function (\ref{hinge}) in (\ref{basis-expansion}) with the generalized hinge function \cite{Wang2005},
  \begin{equation}\label{ghh}
\max\{0, \ell_1(\mathbf{x}),\ldots, \ell_{k_n}(\mathbf{x})\}
 \end{equation}
 with $k_n \leq n$ and $\ell_{k}(\mathbf{x}), k=1,\ldots,k_n$ are affine.
 It has been proved that the GHH model can represent any continuous PWL functions in any dimension. And it is noted that the  activation function employed in the popular  maxout network is the generalized hinge \cite{Goodfellow2013maxout}. In the training of the HH and GHH model, the structural parameters, like the number of basis functions as well as the number of linear functions in each generalized hinge, can not be changed while optimizing the weights parameter $a_1, \ldots, a_M$.

The model of AHH is a PWL representation which
 can be also seen as a linear combination of basis functions, just like (\ref{basis-expansion}), and the basis function takes the form of
  \begin{equation}\label{ahh-basis}
B_m(\mathbf{x})=\min\{g_1(\mathbf{x}),\ldots, g_{K}(\mathbf{x})\}
 \end{equation}
with 
\[
g_k(\mathbf{\mathbf{x}})=\max\{0, s_k(x_{\upsilon_k}-\beta_k)\}, 
\]
$s_k=\pm 1$, $x_{\upsilon_1}, \ldots, x_{\upsilon_k}$ are indices of the input variables, and $\beta_k \in [0,1)$ are predefined distinguished scalars, $k = 1, \ldots, K$. It is proved in \cite{Xu2009ahh} that the equation (\ref{ahh-basis}) can be obtained by restricting $\ell_k(\mathbf{x})$ in (\ref{ghh}) to be univariate linear. Different from the  HH and GHH models, the AHH model is trained using a recursive partitioning procedure, which is similar to the training of  multivariate adaptive regression splines (MARS). The training of the AHH model includes a forward stepwise and a backward stepwise procedures, in which basis functions are added or deleted according to the greedy strategy. It is noted that the structure of the AHH model is dynamically adjusting during the training process. It has been shown in \cite{Xu2009ahh} that the computational complexity increases exponentially with respect to the number of basis functions, and the increase becomes more abrupt when the dimension grows higher and the sample data size is larger.

\section{Efficient hinging hyperplanes neural network}
In this section, we focus on the construction of the EHH neural network, and its properties such as the approximation ability and interpretability. Besides, the relationship between the  AHH model and the EHH neural network is shown.

\subsection{Network Structure}
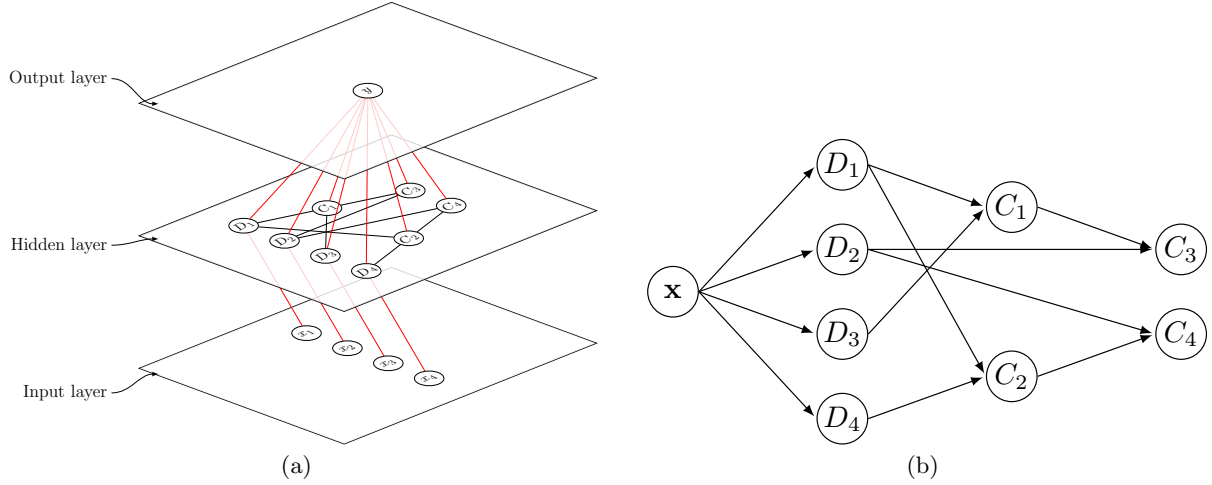
\begin{figure}[htbp]
\centering
\subfigure[]{
    \label{fig_nnstruct} 
    \resizebox{.45\textwidth}{!}{
        \begin{tikzpicture}[scale=.9,every node/.style={minimum size=1cm},on grid]
		\def\layersep{2}
		 \tikzstyle{neuron}=[circle,draw,fill=white,minimum size=17pt,inner sep=0pt]
    \begin{scope}[
            yshift=-120,every node/.append style={
            yslant=0.5,xslant=-1},yslant=.4,xslant=-1.3
            ]
        \draw[black,very thick] (0,0) rectangle (8,-5);
        \fill[white,fill opacity=.9] (0,0) rectangle (8,-5);

        \foreach \name / \y in {1,...,4}
            \path node[neuron] (I-\y) at (4 ,-\y ) {$x_\y$};
    \end{scope}

    \begin{scope}[
    	yshift=0,every node/.append style={
    	    yslant=0.5,xslant=-1},yslant=.4,xslant=-1.3
    	             ]
        
        \foreach \name / \y in {1,...,4}
            \coordinate  (H-\name) at (\layersep ,-\y ) ;
    
        \coordinate  (J-1) at (4, -1.5) ;
        \coordinate  (J-2) at (4, -3.5 ) ;
    
        \coordinate  (J-3) at (6, -2) ;
        \coordinate  (J-4) at (6, -3 ) ;
        
    \end{scope}
    	
    \begin{scope}[
    	yshift=120,every node/.append style={
    	    yslant=0.5,xslant=-1},yslant=.4,xslant=-1.3
    	             ]
    	\coordinate  (Y) at (4, -2.5) ;
    \end{scope}
   
    \draw[thick, -latex, red] (I-1) -- (H-1);
    \draw[thick, -latex, red] (I-2) -- (H-2);
    \draw[thick, -latex, red] (I-3) -- (H-3);
    \draw[thick, -latex, red] (I-4) -- (H-4);

    \begin{scope}[
    	yshift=0,every node/.append style={
    	    yslant=0.5,xslant=-1},yslant=.4,xslant=-1.3
    	             ]
        \draw[black,very thick] (0,0) rectangle (8,-5);
        \fill[white,fill opacity=.8] (0,0) rectangle (8,-5);
    	\foreach \name / \y in {1,...,4}
            \path node[neuron] (H-\name) at (\layersep ,-\y ) {$D_\y$};
    
        \path node[neuron] (J-1) at (4, -1.5) {$C_1$};
        \path node[neuron] (J-2) at (4, -3.5 ) {$C_2$};
    
        \path node[neuron] (J-3) at (6, -2) {$C_3$};
        \path node[neuron] (J-4) at (6, -3 ) {$C_4$};
        
        \path (H-1) edge (J-1);
        \path (H-1) edge (J-2);
        
        \path (H-2) edge (J-3);
        \path (H-2) edge (J-4);
        
        \path (H-3) edge (J-1);
        
        \path (H-4) edge (J-2);
        
        \path (J-1) edge (J-3);
        
        \path (J-2) edge (J-4);
    \end{scope}

    \draw[thick, -latex, red] (J-1) -- (Y);
    \draw[thick, -latex, red] (J-2) -- (Y);
    \draw[thick, -latex, red] (J-3) -- (Y);
    \draw[thick, -latex, red] (J-4) -- (Y);
    
    \draw[thick, -latex, red] (H-1) -- (Y);
    \draw[thick, -latex, red] (H-2) -- (Y);
    \draw[thick, -latex, red] (H-3) -- (Y);
    \draw[thick, -latex, red] (H-4) -- (Y);
    
    \begin{scope}[
    	yshift=120,every node/.append style={
    	yslant=0.5,xslant=-1},yslant=.4,xslant=-1.3
    	             ]
    	\draw[black,very thick] (0,0) rectangle (8,-5);
    	\fill[white,fill opacity=.8] (0,0) rectangle (8,-5);
    	
    	\path node[neuron] (Y) at (4, -2.5) {$y$};
    \end{scope}
    
    \draw[-latex,thick](-.9,5) node[right, anchor=east]{{\large Output layer}}
        to[out=0,in=180] (.6,4.2);
    \draw[-latex,thick](-.9,-.3)node[right, anchor=east]{{\large Hidden layer}}
        to[out=0,in=180] (.6,-.1);
    \draw[-latex,thick](-0.9,-5)node[right, anchor=east]{{\large Input layer}}
        to[out=0,in=180] (.6,-4.4);
\end{tikzpicture}}
    }
\subfigure[]{
    \label{fig_dag} 
        \resizebox{.45\textwidth}{!}{
        \def\layersep{2}

\begin{tikzpicture}[shorten >=1pt,->,draw=black, node distance=\layersep]
    \tikzstyle{every pin edge}=[<-,shorten <=1pt]
    \tikzstyle{neuron}=[circle,draw,fill=white,minimum size=17pt,inner sep=0pt]

        \node[neuron] (I) at (0,-2.5) {$\mathbf{x}$};

    \foreach \name / \y in {1,...,4}
        \path
            node[neuron] (H-\name) at (\layersep cm,-\y cm) {$D_\y$};

    \foreach \dest in {1,...,4}
        \draw[-latex] (I.east) -- (H-\dest.west);

    \path node[neuron] (J-1) at (4cm, -1.5cm) {$C_1$};
    \path node[neuron] (J-2) at (4cm, -3.5 cm) {$C_2$};

    \path node[neuron] (J-3) at (6cm, -2cm) {$C_3$};
    \path node[neuron] (J-4) at (6cm, -3 cm) {$C_4$};
    
    \draw[-latex] (H-1.east) -- (J-1.west);
    \draw[-latex] (H-1.east) -- (J-2.west);
    
    \draw[-latex] (H-2.east) -- (J-3.west);
    \draw[-latex] (H-2.east) -- (J-4.west);
    
    \draw[-latex] (H-3.east) -- (J-1.west);
    
    \draw[-latex] (H-4.east) -- (J-2.west);
    
    \draw[-latex] (J-1.east) -- (J-3.west);
    
    \draw[-latex] (J-2.east) -- (J-4.west);
\end{tikzpicture}
        }
    }
\caption{Illustration of the EHH neural network. (a) The structure of a typical EHH network. (b) Local view of the hidden layer: DAG.}
\label{figure_ehh_sample}
\end{figure}

Fig. \ref{fig_nnstruct} shows a typical EHH neural network. It can be seen as a single hidden layer PWL neural network, which includes an input layer, a hidden layer and an output layer. In the following discussions, we will describe the connections of each 2 subsequent layers. Besides, the hidden layer is described using a directed acyclic graph (DAG), and the connections of nodes in the DAG is explained in detail.

\subsubsection{Pre-processing}

Generally speaking, the input variables for the neural networks may differ from each other in the  ranges of candidate values. For example, the variables may involve the temperature and the humidity  of the environment, which are not in the same units and range of possible values. Thus, in order to avoid computation deficiency and severe variations in the parameter space, we are supposed to pre-process the input data before sending to the hidden layer. Specifically, assume the sampled data is $(\tilde{\mathbf{x}}(k), y(k))_{k=1}^{N_s}$, this is done by normalizing each of the original input variables independently, i.e., 
\[
x_i(k) = \frac{\tilde{x}_i(k)-\min(\tilde{\mathbf{x}}_i)}{\max(\tilde{\mathbf{x}}_i)-\min(\tilde{\mathbf{x}}_i)},
\]
in which $\tilde{\mathbf{x}}_i = [\tilde{x}_i(1), \ldots, \tilde{x}_i(N_s)]^T$. After the standardization process,  each $x_i(k) \in [0,1], \forall i =1, \ldots, n, k=1, \ldots, N_s$. 
 Hereafter in the paper, the input variables refer to be normalized.

\subsubsection{Input and hidden layer connection}
 
 Let $\mathbf{x}=[x_1, \ldots, x_n]^T$ be the pre-processed input variables. Since the hidden layer is essentially a DAG, we define two kinds of nodes (neurons) in the hidden layer, i.e., the source nodes and the intermediate nodes. The source nodes are the nodes that only accept inputs from the input layer, meaning that there are edges directed from the input nodes to the source nodes. Other than the source nodes, we define the intermediate nodes, which can accept inputs from the source nodes  or other intermediate nodes in the hidden layer, i.e., there are edges directed from other nodes to the intermediate nodes.
 
 In the EHH neural network, the source nodes in the hidden layer are designed to have the ``$\max$'' activation function on a single variable in the inputs.  Therefore, the output of a source node can be expressed as
\begin{equation*}
    \mathrm{nn}(\mathbf{x}) = \max\{0, x_{v} - \beta_{v}\},
\end{equation*}
where $v$ is the index of the input variable, and $\beta_{v}$ is the offset. As is shown in Fig. \ref{fig_dag}, the nodes $\mathrm{D}_1, \ldots, \mathrm{D}_{N_d}$ ($N_d=4$) are the source nodes. 

In general, since the EHH neural network employs the univariate activation function, we can express the output of  the source nodes as follows,
 \begin{equation}\label{source}
 \begin{array}{llll}
\mathrm{nn}_{D_1}(\mathbf{x})=\max\{0, x_1\}, & \mathrm{nn}_{D_2}(\mathbf{x})=\max\{0, x_1-\beta_{11}\}, & \ldots, & \mathrm{nn}_{D_q}(\mathbf{x})=\max\{0, x_1-\beta_{1,q-1}\},\\
\vdots & & &\\
\mathrm{nn}_{D_{(n-1)q+1}}(\mathbf{x})=\max\{0, x_n\}, & \mathrm{nn}_{D_{(n-1)q+2}}(\mathbf{x})=\max\{0, x_n-\beta_{n1}\}, & \ldots, & \mathrm{nn}_{D_{nq}}(\mathbf{x})=\max\{0, x_n-\beta_{n,q-1}\}
 \end{array}
 \end{equation}
 in which $\beta_{i1}, \ldots, \beta_{i, q-1}$ are the predefined bias in each dimension. For example, we can take $\beta_{i1}, \ldots, \beta_{i, q-1}$ as the $q$-quantiles in the $i$-th dimension. Hence in total we have $N_d = nq$ source nodes.

\subsubsection{Connection in the hidden layer}
Besides the source nodes, nodes in the hidden layer that are not directly connected to the nodes in the input layer are called the intermediate nodes, such as the node $C_j$, $j = 1, \ldots, N_c$, in Fig. \ref{fig_dag}. In the hidden layer of the EHH neural network, the  intermediate nodes are generated and ordered sequentially, and are designed to take inputs from exactly two other nodes in the hidden layer, and take the  ``$\min$'' activation function. 

In order to construct the EHH neural network systematically, we establish the connection in the hidden layer involving intermediate nodes as follows.

\begin{assum}
Given three nodes $A_{j_1}, A_{j_2}, A_{j_3}$ in the hidden layer in an EHH neural network, if there are edges directed from $A_{j_1}$ and $A_{j_2}$ to $A_{j_3}$, respectively, then the following two rules should be followed, 
\begin{description}
\item[Rule 1] $A_{j_1}, A_{j_2} \in \mathcal{J} \triangleq \{ D_p\mid p = 1, \ldots, N_d \} \cup \{ C_{t} \mid t < j_3 \}$; 

\item[Rule 2] $\mathcal{S}_{A_{j_1}} \cap \mathcal{S}_{A_{j_2}} = \emptyset$, where $\mathcal{S}_{A_{j_1}}$ and $\mathcal{S}_{A_{j_2}}$ are the sets of indices of input variables that appeared in the nodes $A_{j_1}$ and $A_{j_2}$, respectively.
 \end{description}
\end{assum}

  In this way, we can express  the output of $A_{j_3}$ as  
 \begin{equation}\label{nn_connection}
 \mathrm{nn}_{A_{j_3}}(\mathbf{x})=\min\{\mathrm{nn}_{A_{j_1}}(\mathbf{x}), \mathrm{nn}_{A_{j_2}}(\mathbf{x})\}.
 \end{equation}
 
 In order to depict the connections in the hidden layer of an EHH neural network, we introduce a new concept as follows.
 
\begin{definition}
Given an EHH neural network that has $M$ neurons in the hidden layer, the matrix $ \mathbf{A_d} = [\mathbf{A_d}]_{ij} \in \mathbb{R}^{M \times M}$ is called the \textbf{ adjacency matrix} of the EHH neural network, if
\begin{itemize}
\item $\mathbf{A_d}_{ij} = 1$ indicates that there is an edge directed from the $i$-th node to the $j$-th node in the hidden layer, and $\mathbf{A_d}_{ij} = 0$ indicates otherwise.
\end{itemize}
\end{definition}

Specifically, we take the EHH neural network shown in Fig. \ref{fig_dag} for example. Suppos that the source nodes $D_1, \ldots, D_4$  contain different input variables. Without loss of generality, we can denote the output of the four sources nodes as, 
 \begin{equation}\label{ex_ss_node}
\mathrm{nn}_{D_i}(\mathbf{x})=\max\{0, x_{i}-\beta_{i}\}, i = 1, 2, 3, 4.
 \end{equation}

According to the connections involving the intermediate nodes $C_1, \ldots, C_{4}$ in Fig.\ref{fig_dag},  we have 
\[
\begin{array}{l}
\mathrm{nn}_{C_1}(\mathbf{x})=\min\{\mathrm{nn}_{D_1}(\mathbf{x}), \mathrm{nn}_{D_3}(\mathbf{x})\}, \mathrm{nn}_{C_2}(\mathbf{x})=\min\{\mathrm{nn}_{D_1}(\mathbf{x}), \mathrm{nn}_{D_4}(\mathbf{x})\}\\
\mathrm{nn}_{C_3}(\mathbf{x})=\min\{\mathrm{nn}_{D_2}(\mathbf{x}), \mathrm{nn}_{C_1}(\mathbf{x})\}, \mathrm{nn}_{C_4}(\mathbf{x})=\min\{\mathrm{nn}_{D_2}(\mathbf{x}), \mathrm{nn}_{C_2}(\mathbf{x})\}.\\
\end{array}
\]

The index set for each hidden nodes can be written as
\begin{equation}\label{layer1}
\begin{array}{ll}
\mathcal{S}_{D_1}=\{1\}, ~~~\mathcal{S}_{D_2}=\{2\}, &  \mathcal{S}_{D_3}=\{3\},~~~ \mathcal{S}_{D_4}=\{4\},\\
\mathcal{S}_{C_1}=\mathcal{S}_{D_1} \cup \mathcal{S}_{D_3} = \{1, 3\}, & \mathcal{S}_{C_2}=\mathcal{S}_{D_1} \cup \mathcal{S}_{D_4} = \{1, 4\},\\
\mathcal{S}_{C_3}=\mathcal{S}_{D_2} \cup \mathcal{S}_{C_1}=\{1, 2, 3\}, & \mathcal{S}_{C_4}=\mathcal{S}_{D_2}\cup \mathcal{S}_{C_2}=\{1,2,4\}.
\end{array}
\end{equation}

Then the adjacency matrix ${\mathbf{A_d}}$ that describes the connections of nodes in the DAG for the EHH network shown in Fig.\ref{fig_nnstruct} can be written as
 \begin{equation}\label{adja_ex}
 \mathbf{A_d}= \left[
 \begin{array}{r|cccc:cc:cc}
 {}&D_1&D_2&D_3&D_4&C_{1}&C_{2}&C_{3}&C_{4}\\
 \hline
 D_{1}&0&0&0&0&1&1&0&0\\
 D_{2}&0&0&0&0&0&0&1&1\\
 D_{3}&0&0&0&0&1&0&0&0\\
 D_{4}&0&0&0&0&0&1&0&0\\
  \hdashline
 C_{1}&0&0&0&0&0&0&1&0\\
 C_{2}&0&0&0&0&0&0&0&1\\
 \hdashline
 C_{3}&0&0&0&0&0&0&0&0\\
 C_{4}&0&0&0&0&0&0&0&0
 \end{array}\right]
 \end{equation}


The following remarks give further insights into the construction of the connections among the neurons in the hidden layer of the EHH neural network.

\begin{remark}
There should be only two ``1"s in a column in $\mathbf{A_d}$ as each neuron has exactly two inputs as is designed, and there may be multiple ``1"s in a row, as each neuron may be inputs to more than 1 neuron. 
\end{remark}

\begin{remark}
The input variables involved in each term of a single neuron are ensured to be distinct, which guarantees that none of the terms can be removed without changing the output of the neuron, i.e., the parameters in the neuron are irredundant. For example, without loss of generality, if we include in one neuron two identical input variables, such as,
\begin{equation*}
\mathrm{nn}_{C_1}(\mathbf{x})=\min\{\max\{0, x_i-\beta_{k_i}\}, \max\{0, x_i-\beta_{k_j}\}\},
\end{equation*}
where $\beta_{k_i} < \beta_{k_j}$, we can rewrite $\mathrm{nn}_{C_1}(\mathbf{x})$ as
\begin{equation*}
\mathrm{nn}_{C_1^\prime}(\mathbf{x})=\min\{\max\{0, x_i-\beta_{k_j}\}\},
\end{equation*}
which implies that the parameters in the neuron $C_1$ are redundant and can be simplified to the neuron $C_1^\prime$. Therefore, the number of input variables contained in each neuron will not exceed the input dimension, $n$.

\end{remark}

\subsubsection{Hidden and output layer Connection}

In the EHH neural network, both the source and intermediate nodes are designed to be connected to the nodes in the output layer, as Fig. \ref{fig_nnstruct} shows. Therefore, the output of the EHH neural network takes the form of the weighted sum of all neurons (with a bias) in the hidden layer, i.e.,
\begin{equation}\label{output}
{f}_{\mathrm{EHH}}(\mathbf{x})=\sum\limits_{k=1}^{M}\alpha_{k}\mathrm{nn}_{A_k}(\mathbf{x})+\alpha_0,
\end{equation}
in which $A_k$ is the neuron in the hidden layer, $\mathrm{nn}_{A_k}(\mathbf{x})$ and $\alpha_{k} \in \mathbb{R}$ denote the output and coefficient of the neuron $A_k$, $k \in  \{1, \ldots, M\}$, and $\alpha_0$ is the constant bias. The coefficients $\alpha_0, \alpha_1, \ldots, \alpha_M$ are called \textbf{weights of the EHH neural network}.

 \begin{remark}
When the EHH neural network is configured to have $N_d = n$ source nodes and $N_c = 0$ intemediate nodes, and the offsets $\beta_i$ are set to be $0$, $\forall i$, we have
\[
\mathrm{nn}_{D_i}(\mathbf{x})=x_i, i = 1, \ldots, n,
\]
since $\mathbf{x} \in [0, 1]^n$, and the EHH neural network is degenerated to a linear approximator,
\begin{equation}
f_{\mathrm{EHH}}(\mathbf{x})=\alpha_0+\sum\limits_{i=1}^n \alpha_i x_i.
\end{equation}
 \end{remark}


\subsection{EHH neural network versus AHH model}

The EHH neural networks are also within the HH family, in particular, we will explain that the EHH neural network can find the equivalent AHH model. It has been shown in \cite{Xu2009ahh} that the AHH model is a tree structure, which is obtained by recursively partitioning of the domain coordinatewisely. Each partition of the domain (say $x_{\upsilon_i}$ at $\beta_{k_{\upsilon_i}}$) generates 2 new bases by taking the minimum of the current basis functions $B_1,\ldots, B_M$ and $\max\{0, x_{\upsilon_i}-\beta_{k_ {\upsilon_i}}\}$ or $\max\{0, -(x_{\upsilon_i}-\beta_{k_{\upsilon_i}})\}$, i.e., $\min\{B_1,\ldots, B_M, \max\{0, x_{\upsilon_i}-\beta_{k_{\upsilon_i}}\}\}$ or $\min\{B_1,\ldots, B_M, \max\{0, -(x_{\upsilon_i}-\beta_{k_{\upsilon_i}})\}\}$. Then the backward stepwise procedure deletes redundant basis functions according to a greedy strategy. 

\begin{figure}[htbp]
  \centering
     \includegraphics[width=0.45\columnwidth]{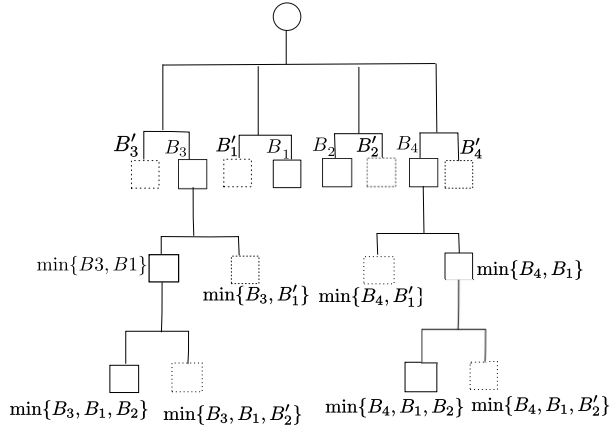}
       \caption{A simple AHH.}
  \label{fig-ahh-ehh-equ}
\end{figure}

We start from a simple example. Fig. \ref{fig-ahh-ehh-equ} is a simple AHH tree structure, in which the circle denotes the root, and the squares denote the leaves. The first generation takes the form of $B_1=\max\{0, x_{1}-\beta_{1}\}, B_1^{\prime} = \max\{0, -(x_{1}-\beta_{1})\}, B_2=\max\{0, x_{2}-\beta_{{2}}\}, B_2^{\prime}=\max\{0, -(x_{2}-\beta_{{2}})\}, B_3=\max\{0, x_{3}-\beta_{{3}}\}, B_3^{\prime}=\max\{0, -(x_{3}-\beta_{{3}})\}$ and $B_4=\max\{0, x_{4}-\beta_{{4}}\}, B_4^{\prime}=\max\{0, -(x_{4}-\beta_{{4}})\}$. 

It can be seen in Fig. \ref{fig-ahh-ehh-equ} that the first generation of the AHH tree, i.e., $B_1, B_1^{\prime}, B_2, B_2^{\prime}, B_3, B_3^{\prime}, B_4, B_4^{\prime}$ are derived by splitting the domain at $x_{1}=\beta_{1}, x_{2}=\beta_{{2}}$, $x_{3}=\beta_{ {3}}, x_{4}=\beta_{ {4}}$, respectively. 
The second generation of the AHH tree is obtained by partitioning the regions in which the basis function $B_3$ and $B_4$ are active ($x_{3} \geq \beta_{{3}}$ and $x_{4} \geq \beta_{{4}}$ respectively) at $x_{1}=\beta_{{1}}$. Then for the third generation, the splitting occurs at $x_2=\beta_{2}$. Suppose in this case the leaves containing terms $\max\{0, -(x_i-\beta_{i})\}, i=1,2,3,4$ are all removed during the backward stepwise process.

In the AHH tree structure, the relationships between the fathers and daughters are clear. 
However, different daughters may contain the genes from the same father, say the daughters $\min\{B_3,B_1\}$ and $\min\{B_4,B_1\}$. And in the tree structure, no connections indicate this. Besides, this representation is not a distributed representation. For example, the calculation of $\min\{B_3, B_1\}$ does not use the information calculated for $B_1$, hence for this simple tree structure, the information of $B_1$ has to be calculated for 3 times.

It is not hard to verify the AHH tree shown in Fig. \ref{fig-ahh-ehh-equ} and the EHH hidden layer shown in Fig. \ref{fig_dag} are equivalent (Let $\mathrm{nn}_{D_i}$ correspond to $B_i, i=1, 2, 3, 4$, respectively). From Fig. \ref{fig-ahh-ehh-equ}, we can see that the first generation of the AHH tree corresponds to the first column in the DAG.
In the EHH network hidden layer, the interaction between $B_1$ and $B_3$ ($\mathrm{nn}_{D_1}$ and $\mathrm{nn}_{D_3}$) is shown clearly, thus the information of $B_1$ and $B_3$ can be reused. This is especially useful when the network outputs need to be calculated many times, hence the EHH networks in Fig. \ref{fig_nnstruct} is a distributed representation. 

It is noted that in the AHH tree, the term $\max\{0, -(x_{\upsilon}-\beta_{k_{ \upsilon}})\}$ exists in the bases, while in the EHH neural network, the sign ``-'' does not appear. The following lemma explains the relationship of the 2 representations.

\begin{lem}
Every EHH neural network can find an equivalent AHH tree.
\end{lem}
\begin{pf}
In the EHH networks, assume a node $A_{j}$ receives inputs from the neurons $A_{j_1}$, $A_{j_2}$, i.e., 
\[\mathrm{nn}_{A_j}=\min\{\mathrm{nn}_{A_{j_1}}, \mathrm{nn}_{A_{j_2}}\}.
\]
 The same rule applies to $\mathrm{nn}_{A_{j_1}}$ and $\mathrm{nn}_{A_{j_2}}$, thus finally we can rewrite $\mathrm{nn}_{A_j}$ as
\begin{equation}\label{exp_fkm}
\mathrm{nn}_{A_j}=\min_{k \in \mathcal{K}_j}\{\mathrm{nn}_{A_k}\}
\end{equation}
where $\mathcal{K}_j$ are the index sets of the source nodes that have direct or indirect connections to $\mathrm{nn}_{j}$.  Compared with the AHH basis function in \cite{Xu2009ahh}, which can be written as
\begin{equation}
B_m(\mathbf{x})=\min\limits_{k \in \mathcal{K}_m}\{\max\{0, s_{km}(x_{\upsilon_{km}}-\beta_{km})\}\},
\end{equation}
with $s_{km}=\pm 1$. By letting all the $s_{km}$ to be 1, we know that the outputs of the EHH nodes can be written as equivalent AHH basis functions. As each node can find an equivalent AHH basis, every EHH neural network can find an equivalent AHH tree.
\end{pf} 
However, for an AHH tree, as the possible existence of $\max\{0, -(x_{i}-\beta_{k, i})\}$, there may be no equivalent EHH network. As the model of AHH can approximate any continuous functions in a compact domain, attention is paid to the approximation ability of the EHH neural networks.

\subsection{Approximation ability of EHH neural networks}
 
 For the case of simplicity, we consider the approximation on the compact set $[0,1]^n$.

\begin{thm}\label{appro-lemma}
Let $f: [0,1]^n \rightarrow \mathbb{R}$ be a continuous function. Then for any $\varepsilon>0$, there exists an EHH network $f_{\rm{EHH}}$, such that for all $\mathbf{x} \in [0,1]^n$, we have
\begin{equation}\label{ehh-appro-thm}
\|f(\mathbf{x})-f_{\rm{EHH}}(\mathbf{x})\|<\varepsilon.
\end{equation}
\end{thm}

\begin{pf}
Assume the source nodes of the EHH neural network are
\begin{equation*}
\max\{0, x_1-\beta_{1}\}, \ldots, \max\{0, x_1-\beta_q\}, \ldots, \max\{x_n-\beta_1\}, \ldots, \max\{x_n-\beta_{q}\}
\end{equation*}
with $\beta_1=0, \beta_2=1/q, \ldots, \beta_q=(q-1)/q$, in which $q$ is a positive integer and can be tuned.

For the intermediate nodes, the full connection strategy is employed, i.e., taking the minimum of any 2 existing neurons with different input variables to form a new neuron. Notice that this full connection strategy is different from ``real'' full connection, as some connections are duplicated or contradict Rule 2. 

The resulting EHH neural network can be written as (arranged according to the number of input variables contained in a neuron)
\begin{align}\label{HL-CPWL}
f_{\mathrm{EHH}}(\mathbf{x})&=\alpha_0+\sum\limits_{k_1=1}^{n}\sum\limits_{j_1=1}^{q}{a_{j_1}^{(k_1)}\max\left\{0,x_{k_1}-\beta_{j_1}\right\}} \nonumber\\
&+\sum\limits_{k_1=1}^{n-1}\sum\limits_{k_2=k_1+1}^{n}\sum\limits_{j_1=1}^{q}\sum\limits_{j_2=1}^{q}a_{j_1j_2}^{(k_1k_2)}\cdot\min\left\{\max\{0, x_{k_1}-\beta_{j_1}\},\max\left\{0, x_{k_2}-\beta_{j_2}\right\}\right\}\nonumber\\
&+\cdots+\sum\limits_{k_1=1}^{n-r+1}\cdots\sum\limits_{k_r=k_1+r-1}^{n}\sum\limits_{j_1=1}^{q}\cdots
\sum\limits_{j_r=1}^{q}a_{j_1\ldots j_r}^{(k_1 \ldots
k_r)}\cdot \min\left\{\max\{0, x_{k_1}-\beta_{j_1}\}, \cdots, \max\left\{0,x_{k_r}-\beta_{j_r}\right\}\right\}\nonumber\\
&+\cdots +\sum\limits_{j_1=1}^{q}\cdots
\sum\limits_{j_n=1}^{q}a_{j_1\ldots
 j_n}\cdot\min\left\{\max\{0,x_1-\beta_{j_1}\}, \cdots,\max\left\{0,x_{n}-\beta_{j_n}\right\}\right\}.
\end{align}
The number of neurons (plus the constant neuron) is 
\begin{equation}\label{eqn_number}
1+\left(_1^n\right)\cdot q+\left(_2^n\right)\cdot
q^2+\cdots+\left(_r^n\right)\cdot q^r+\cdots+\left(_n^n\right)q^n=(q+1)^n.
\end{equation}
 
 Fig. \ref{fig-full-connection} illustrates the full connection strategy in the EHH neural network with $n=3$ and $q=2$, in which the neurons including the same number of input variables are arranged in the same column. The source nodes $D_1, D_2$ involve the input variable $x_1$, while $D_3, D_4$ contain $x_2$ and $D_5, D_6$ consist of $x_3$, respectively.
 
 It is noted that according to (\ref{eqn_number}), there are supposed to be $23$ intermediate nodes in the network. However, only $20$ intermediate nodes are generated in Fig. \ref{fig-full-connection}. One reason is that the neurons containing the same variables, e.g. $D_1$ and $D_2$,  are not allowed to form a new neuron according to Rule 2.  Another reason is that  some combinations may be duplicate. For example, $\min\{\mathrm{nn}_{D_1}, \mathrm{nn}_{C_9}\}$, $\min\{\mathrm{nn}_{D_3}, \mathrm{nn}_{C_3}\}$ are equivalent (all equal to $\min\{\mathrm{nn}_{D_1}, \mathrm{nn}_{D_3}, \mathrm{nn}_{D_5}\}$), and we can only keep 1 combination to formulate a new neuron $C_{13}$.

\begin{figure}[htbp]
\centering
\begin{tikzpicture}[shorten >=1pt,->,draw=black, node distance=\layersep]
    \tikzstyle{every pin edge}=[<-,shorten <=1pt]
    \tikzstyle{neuron}=[circle,draw,fill=white,minimum size=17pt,inner sep=0pt]

    \foreach \name / \y in {1,...,6}
        \path node[neuron] (H-\name) at (2 , -\y) {$D_\y$};

    \foreach \name / \y in {1,...,12}
        \pgfmathparse{-\y+3.3}
        \path node[neuron] (J-\name) at (8 ,\pgfmathresult  ) {$C_{\y}$};

    \foreach \name / \y in {13,...,20}
        \pgfmathparse{-\y*1.2+15.6}
        \path node[neuron] (J-\name) at (14 ,\pgfmathresult  ) {$C_{\y}$};
        
    \foreach \name / \y in {1,2,3,4}
        \draw[-latex] (H-1.east)  --  (J-\y.west);
    
    \foreach \name / \y in {5,6,7,8}
        \draw[-latex] (H-2.east)  --  (J-\y.west);
    
    \foreach \name / \y in {5,9,1,10}
        \draw[-latex] (H-3.east)  --  (J-\y.west);
        
    \foreach \name / \y in {2,6,11,12}
        \draw[-latex] (H-4.east)  --  (J-\y.west);
        
    \foreach \name / \y in {3,7,9,10}
        \draw[-latex] (H-5.east)  --  (J-\y.west);
        
    \foreach \name / \y in {4,8,10,12}
        \draw[-latex] (H-6.east)  --  (J-\y.west);
        
    \foreach \name / \y in {13,14,15,16}
        \draw[-latex, red] (H-1.east)  --  (J-\y.west);
        
    \foreach \name / \y in {17,18,19,20}
        \draw[-latex, red] (H-2.east)  --  (J-\y.west);
        
    \draw[-latex, red] (J-9.east)  --  (J-13.west);
    \draw[-latex, red] (J-10.east)  --  (J-14.west);
    \draw[-latex, red] (J-11.east)  --  (J-15.west);
    \draw[-latex, red] (J-12.east)  --  (J-16.west);
    \draw[-latex, red] (J-9.east)  --  (J-17.west);
    \draw[-latex, red] (J-10.east)  --  (J-18.west);
    \draw[-latex, red] (J-11.east)  --  (J-19.west);
    \draw[-latex, red] (J-12.east)  --  (J-20.west);
\end{tikzpicture}
         \caption{Illustration of the full connection of neurons in the hidden layer of the EHH network with $n = 3$ and $q = 2$. The neurons with the same number of input variables are arranged in the same column, and $D$ and $C$ stand for the source and intermediate nodes, respectively.}
  \label{fig-full-connection}
\end{figure}
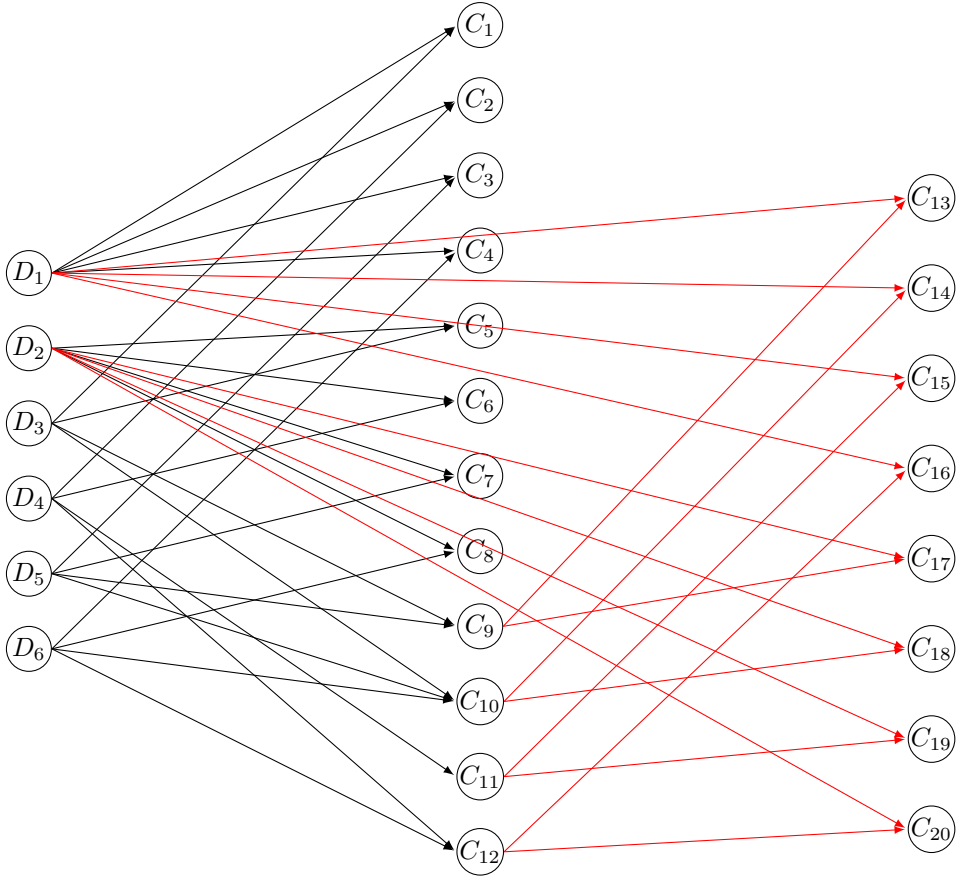
  
It has been proved in \cite{Julian1999} that the boundary configuration
defined by (\ref{HL-CPWL}) subdivides the hypercube $[0,1]^n$ into the simplices with a scaling factor $1/q$. Letting the equation (\ref{HL-CPWL}) equal $f(x)$ at the vertex of the
simplicies yields unique coefficients $a_{j_1\ldots j_r}^{(k_1\ldots k_r)}(r = 1,\ldots, n)$. Moreover, the following holds,
\begin{equation*}
\|f_{\mathrm{EHH}}(\mathbf{x})-f(\mathbf{x})\| \leq \epsilon, \forall \mathbf{x} \in [0,1]^n
\end{equation*}
where $\epsilon=\max\limits_{\Delta \in [0,1]^n}\max\limits_{\mathbf{x}_1, \mathbf{x}_2 \in \Delta}\|f(\mathbf{x}_1)-f(\mathbf{x}_2)\|$, and $\Delta$ denotes the simplices.

As for any given $\varepsilon$, due to the continuity of $f$ on the compact set $[0,1]^n$, we can find $\delta$ such that
\begin{equation*}
\|f(\mathbf{x}_1)-f(\mathbf{x}_2)\| < \varepsilon, ~~\forall \|\mathbf{x}_1-\mathbf{x}_2\|<\delta.
\end{equation*}
Choosing $q$ to ensure 
\begin{equation*}
\|\mathbf{x}_1-\mathbf{x}_2\|<\delta, \forall \mathbf{x}_1, \mathbf{x}_2 \in \Delta, \forall \Delta \in [0,1]^n,
\end{equation*}
then we have
\begin{equation*}
\epsilon=\max\limits_{\Delta \in [0,1]^n}\max\limits_{\mathbf{x}_1, \mathbf{x}_2 \in \Delta}\|f(\mathbf{x}_1)-f(\mathbf{x}_2)\|<\varepsilon.
\end{equation*}

Therefore the conclusion (\ref{ehh-appro-thm}) follows.

\end{pf}

As is shown in the proof of Theorem \ref{appro-lemma}, the EHH neural network can approximate any continuous function in a compact domain with arbitrary precision, if sufficiently large number of neurons are provided.  In other words, it requires that the simplices $\Delta$ are small enough.  Actually, in the simulation study, we will show that the number of the neurons will not exceed a few hundreds in approximating a complex function in a few tens dimension, and the training of the EHH neural network will remove some of the preset neurons.

\subsection{Interpretability of EHH}
\subsubsection{Interaction matrix}

To describe the interpretability of the EHH network, we introduce the \textbf{interaction matrix} $\mathbf{Ir}$ to depict the interactions among different input variables. 

\begin{definition}
Given an EHH neural network that has $M$ neurons in the hidden layer, the matrix $\mathbf{Ir} = [\mathbf{Ir}]_{ij} \in \mathbb{R}^{M \times M}$ is called the \textbf{interaction matrix} of the EHH neural network, if it satisfies 
\begin{itemize}
\item $\mathbf{Ir}_{ij} = 1$ indicates that there is a path (maybe not direct) from the node $A_i$ to the node $A_j$, and $\mathbf{Ir}=0$ indicates otherwise;
\item $\mathbf{Ir}_{ij} = 0$, $\forall j$, if $A_i$ is an intermediate node.
\end{itemize}
\end{definition}

As (\ref{exp_fkm}) indicates, for each neuron $A_j$, we can find the univariate neurons that constitute it, i.e., find the source nodes that have connections to the neuron $A_j$ (may be not direct connection). Although the interaction matrix can not be obviously obtained from the structure of the EHH neural network, it can be obtained from the adjacency matrix, as Algorithm \ref{adja2ir} shows.
\algrrule[0.8pt]
\begin{alg}
{Obtaining the interaction matrix for an EHH neural network.}
\label{adja2ir}
\algrule[0.5pt]
\begin{algorithmic}[1]
\hspace{-4ex} \textbf{Input:} The EHH neural network $f_{\mathrm{EHH}}$ with the adjacency matrix $\mathbf{A_d}$.\\
\hspace{-4ex} \textbf{Output:} The interaction matrix $\bf Ir$ for the EHH neural network.
\STATE Initialize $\mathbf{Ir}=\mathbf{A_d}$;
\FOR{$j =1: M$ }
\STATE Let $\mathcal{I}=\{j\}$.
\FOR{$i \in \mathcal{I}$}
\IF{$i>N_d$}
\STATE Find $i_1, i_2$ such that $\mathbf{Ir}_{i_1,i}=1,\mathbf{Ir}_{i_2,i}=1$, add $i_1, i_2$ to the set $\mathcal{I}$.
\STATE Let $\mathbf{Ir}_{i,j}=0$.
\ELSE
\STATE Let $\mathbf{Ir}_{i,j}=1$.
\ENDIF
\ENDFOR
\ENDFOR
\algrule[0.5pt]
\end{algorithmic}
\end{alg}

In a word, the adjacency matrix ${\mathbf{A_d}}$ represents the connections of each neuron while the interaction matrix $\mathbf{Ir}$ describes the interactions among different input variables. For example, the interaction matrix for Fig. \ref{fig_dag} are listed.
 
 \begin{equation}
\mathbf{Ir}= \left[ \begin{array}{r|cccc:cc:cc}
{}&D_1&D_2&D_3&D_4&C_{1}&C_{2}&C_{3}&C_{4}\\
\hline
 D_{1}&0&0&0&0&1&1&1&1\\
 D_{2}&0&0&0&0&0&0&1&1\\
 D_{3}&0&0&0&0&1&0&1&0\\
 D_{4}&0&0&0&0&0&1&0&1\\
  \hdashline
 C_{1}&0&0&0&0&0&0&0&0\\
 C_{2}&0&0&0&0&0&0&0&0\\
 \hdashline
 C_{3}&0&0&0&0&0&0&0&0\\
 C_{4}&0&0&0&0&0&0&0&0
 \end{array}\right].
 \end{equation}
 
 \begin{remark}
 Both the adjacency matrix and interaction matrix are sparse, which is easy to be stored and applied to large-scale problems.
 \end{remark}
 
\subsubsection{ANOVA decomposition}

Similar to the ANOVA decomposition used in \cite{Friedman1991}, here we could easily get the ANOVA decomposition of the EHH neural network through the interaction matrix $\mathbf{Ir}$,
\begin{equation}\label{anova-ehh}
f_{\mathrm{EHH}}(\mathbf{x})=\alpha_0+\sum\limits_i\sum\limits_{|\mathcal{S}_{A_j}|=1} \alpha_j \mathrm{nn}_{A_j}(x_i)+\sum\limits_{i,k}\sum\limits_{|\mathcal{S}_{A_j}|=2} \alpha_j \mathrm{nn}_{A_j}(x_i, x_k)+\sum\limits_{i,k,r}\sum\limits_{|\mathcal{S}_{A_j}|=3}\alpha_j \mathrm{nn}_{A_j}(x_i, x_k, x_r),
\end{equation}
in which $|\mathcal{S}_{A_j}|$ denotes the number of elements in the set $\mathcal{S}_{A_j}$, i.e., the number of nonzero elements in $\mathbf{Ir}(:, j)$. The first sum is over all source nodes, the second sum is over all neurons having exactly 2 input variables, representing (if present) two-variable interactions. Similarly, the third sum represents (if present) the contributions from three-variable interactions in corresponding neurons and so on. As (\ref{anova-ehh}) is similar to the decompositions provided by the analysis of variance for contingency tables, analog to the definition, we refer to (\ref{anova-ehh}) as the ANOVA decomposition of the EHH neural network.

It is easy to derive (\ref{anova-ehh}) for an EHH neural network. For example, each function in the first sum of (\ref{anova-ehh}) can be expressed as
\begin{equation}\label{uni_ehh}
f_i(x_i)=\sum\limits_{\mathcal{S}_{A_j}=\{i\}} \alpha_{j}\mathrm{nn}_{A_j}(\mathbf{x}),
\end{equation}
which is a sum over all source nodes involving the particular variable $x_i$. 

Each bivariate function in the second sum of (\ref{anova-ehh}) can be expressed as
\begin{equation}\label{bi_ehh}
f_{i,k}(x_i, x_k)=\sum\limits_{\mathcal{S}_{A_j}=\{i, k\}}\alpha_{j}\mathrm{nn}_{A_j}(\mathbf{x}),
\end{equation}
which is a sum over all neurons involving the particular pair of the variables $x_i$ and $x_k$. Adding this to the corresponding univariate contributions (\ref{uni_ehh}) (if present) indicates the joint bivariate contribution of $x_i$ and $x_k$ to the model. Similarly, each trivariate function in the third sum can be obtained by collecting together all neurons involving the particular variable triples, i.e.,
\begin{equation}\label{tri_ehh}
f_{i,k,r}(x_i, x_k, x_r)=\sum\limits_{\mathcal{S}_{A_j}=\{i,k,r\}}\alpha_{j}\mathrm{nn}_{A_j}(\mathbf{x}).
\end{equation}
Adding this to the corresponding univariate and bivariate functions (\ref{uni_ehh}), (\ref{bi_ehh}) involving $x_i, x_k$ and $x_r$, provides the joint contribution of these three variables to the model. Terms involving more variables (if present) can be collected together and represented similarly.

{The summation (\ref{uni_ehh})-(\ref{tri_ehh}) can be easily determined through the interaction matrix}. For example, in order to determine the index $j$ such that $\mathcal{S}_{A_j} = \{i, k, r\}$, 
assume the variables $x_i$, $x_k$ and $x_r$ are contained in the source nodes $A_{\gamma_i}$, $A_{\gamma_k}$ and $A_{\gamma_r}$, respectively. That means $\mathcal{S}_{A_{\gamma_i}}=\{i\}$, $\mathcal{S}_{A_{\gamma_k}}=\{k\}$, and $\mathcal{S}_{A_{\gamma_r}}=\{r\}$. It is noted that such $\gamma_i, \gamma_k$ and $\gamma_r$ may not be unique. Then if for any $\gamma_i, \gamma_k, \gamma_r$, the following is true,
\[
\mathbf{Ir}_{\gamma_i, j}=1,\mathbf{Ir}_{\gamma_k, j}=1,\mathbf{Ir}_{\gamma_r, j}=1,
\]
we have $\mathcal{S}_{A_j}=\{i, k, r\}$.

{Interpretation of the EHH neural network is greatly facilitated through its ANOVA decomposition} (\ref{anova-ehh}). From (\ref{anova-ehh}), we can know the importance of the input variables as well as the interactions among different input variables appeared in the EHH neural network .

\section{Training of the EHH neural network}



For neural networks, suitable initial parameters of the model, like the number of layers and the number of neurons, are essential. To search for a good neural network structure, conventional approaches apply evolution algorithm or reinforcement learning over a discrete and non-differentiable search space, which is computationally demanding \cite{Zoph2018learning}. One kind of speeding up approaches includes imposing a particular structure of the search space \cite{Liu2018hierarchical}. 

Now that the structure of the EHH neural network can be described through the adjacency matrix like (\ref{adja_ex}), 
the determination of the EHH neural network can be formulated as determining the binary parameters in the adjacency matrix as well as continuous variables $\alpha_j, j\in \{0, 1, \ldots, M\}$ in (\ref{output}). 

As is well known, to simultaneously obtain the optimal binary and continuous variables, we have to solve a mixed-integer nonlinear programming problem, which is NP-hard. For the EHH neural networks with large number of neurons, this is prohibitive. Hence in our work, we propose a descent algorithm, which search the solution in a sub-space, but yet result in a local optimal solution. Besides, the worst-case complexity analysis as well as a speeding up strategy of the algorithm is given.

\subsection{Initial EHH neural network generation}
Given the pre-defined number of neurons $M$, the initial EHH network with $M$ neurons is generated randomly according to Rule 1 and 2.  Apart from the source nodes, as each neuron (node) in the hidden layer is generated by taking the minimum of 2 existing neurons, we have
\begin{equation}\label{generation_strategy}
\mathbf{A_d}(i, j)=0, ~\forall i \geq j, j \in \{1, \ldots, M\}.
\end{equation}

In order to determine the weights $\alpha_j, j=0, 1, \ldots, M$ in (\ref{output}), an $\ell$-1 regularization problem, namely Lasso optimization problem is solved. The reason of imposing an $\ell$-1 penalty is to remove neurons that are either dependent on each other or contribute less to the output. Specifically, Lasso pruning is done through the optimization problem described below:
\begin{equation}\label{lasso-prune}
\begin{array}{ll}
\min\limits_{\bm \alpha}&J(\alpha)=(y-\mathbf{Z}\alpha)^T(y-\mathbf{Z}\alpha)+\lambda \|\bm \alpha\|_1,
\end{array}
\end{equation}
where $\bm{\alpha}=[\alpha_0, \alpha_{1}, \ldots, \alpha_{M}]^T$, {$\mathbf{Z}$} is called the data matrix
\begin{equation}\label{data_matrix}
\mathbf{Z}=\left[\begin{array}{cccc}
1&\mathrm{nn}_{1}(\mathbf{x}(1)) & \cdots&\mathrm{nn}_{M}(\mathbf{x}(1))   \\
 \vdots&\vdots&\vdots & \vdots    \\
1&\mathrm{nn}_{1}(\mathbf{x}(N_s))  & \cdots&\mathrm{nn}_{M}(\mathbf{x}(N_s))   
\end{array}
\right].
\end{equation}
The first all one column in $\mathbf{Z}$ represents the constant bias, $\{\mathbf{x}(k),y(k)\}_{k=1}^{N_s}$ are the training samples, and $\mathrm{nn}_j(\mathbf{x}(k))$ is the output of the neuron $A_j$ for the $k$-th sample. 

\begin{remark}
The shrinkage parameter $\lambda$ controls the amount of sparsity, which should be selected according to specific problems. A common choice of $\lambda$ is done through cross validation. Here we assume
\begin{equation}\label{lambda-choosing}
\lambda=\zeta \sqrt{2\log(l_{\alpha})},
\end{equation}
where $l_{\alpha}$ denotes the length of $\bm \alpha$. 

The parameter $\zeta$ is chosen from a set of values (different for specific problems) as the one yielding the least generalized cross-validation criterion GCV \cite{Craven1979},
\begin{equation}\label{gcv}
\mathrm{GCV}(M)=\frac{\sum\limits_{i=1}^{N_s} [y(i)-f_{\mathrm{EHH}}(\mathbf{x}(i))]^2}{{N_s}\left[1-\frac{C(M)}{N_s}\right]^2},
\end{equation}
in which the complexity function takes the form of
\begin{equation}
C(M)=\mathrm{trace}(\mathbf{Z}(\mathbf{Z}^T\mathbf{Z})^{-1}\mathbf{Z}^T)+d\cdot M
\end{equation}
according to \cite{Friedman1991}. The quantity $d$ represents a cost for each neuron  optimization and larger value for $d$ will lead to fewer neurons in the network. The principle of the selection of $d$ can be found in \cite{Friedman1991}, and accordingly, here we choose $d=3$. The term $\mathrm{trace}(\mathbf{Z}(\mathbf{Z}^T\mathbf{Z})^{-1}\mathbf{Z}^T)$ is actually the column rank of the matrix $\mathbf{Z}$. When the columns of $\mathbf{Z}$ are independent, i.e., the outputs of all neurons are independent, we have $C(M)=M+1$. 

\end{remark}

By solving the problem (\ref{lasso-prune}), we have introduced sparsity in the entries of $\bm{\alpha}$, i.e., the connections of some neurons to the output become 0, meaning that such neurons contribute much less significantly to the output. Introducing an augmented adjacency matrix $\mathbf{\bar{A}_d}$,
\begin{equation}\label{augmented_ad}
\mathbf{\bar{A}_d}=\left[\begin{array}{cc}{}&\alpha_1\\
\mathbf{A_d}&\vdots\\
{}&\alpha_M
\end{array}\right],
\end{equation}
then there may be rows with all zero elements in ${\mathbf{\bar{A}_d}}$, i.e., there may be neurons that are neither connected to other neurons nor to the output. We can delete these neurons in the network without affecting the performance of the EHH neural network. This can be fulfilled through deleting corresponding rows and columns in the augmented adjacency matrix ${\mathbf{\bar{A}_d}}$, and such deletion can be repeated until there are no all zero rows. Take Fig. \ref{fig_dag} for example, the augmented adjacency matrix is,
  \begin{equation}\label{eqn_delete}
 \mathbf{\bar{A}_d}= \left[
 \begin{array}{c|ccccccccc}
 {}&D_1&D_2&D_3&D_4&C_{1}&C_{2}&C_3&C_4\makebox(-11,0){\rule[-65ex]{0.4pt}{14\normalbaselineskip}} &y\\
 \hline
 D_{1}&0&0&0&0&1&1&0&0&\alpha_1\\
 D_{2}&0&0&0&0&0&0&1&1&\alpha_2\\
 D_{3}&0&0&0&0&1&0&0&0&\alpha_3\\
 D_{4}&0&0&0&0&0&1&0&0&\alpha_4\\
 C_{1}&0&0&0&0&0&0&1&0&\alpha_5\\
 C_{2}&0&0&0&0&0&0&0&1&\alpha_6 \\
 C_{3}&0&0&0&0&0&0&0&0&\alpha_7\\
 C_{4} &0&0&0&0&0&0&0&0&\alpha_8 \makebox(-150,0){\rule[1.5ex]{15\normalbaselineskip}{0.4pt}}
 \end{array}\right].
 \end{equation}

If $\alpha_8$ becomes 0 after solving the optimization problem (\ref{lasso-prune}), which means $C_4$ is not connected to the output, then both the 8-th row and 8-th column can be deleted in $\mathbf{\bar{A}_d}$. Then if further $\alpha_6=0$,  as $C_4$ has been removed, the neuron $C_2$ is not connected to any other neurons nor the output, both the 6-th row and 6-th column can be also deleted. Notice that if $\alpha_6=0, \alpha_8 \neq 0$, no neurons can be deleted, as there are no all zero rows in this case, which means that although $C_2$ is not connected to the output, it has connection to $C_4$. 
After deleting all redundant neurons,  the neural network is much more condensed and the new adjacency matrix $\mathbf{A_d}$ can be obtained by just removing the last column in the resulting $\mathbf{\bar{A}_d}$.


\subsection{Optimization of the EHH neural network}
The optimization of the EHH neural network refers to the simultaneously determining the continuous and binary variables, and can be cast as the problem below:
\begin{equation}\label{EHH-optimization}
\begin{array}{ll}
\min\limits_{\mathbf{A_d}, \bm{\alpha}}&J(\mathbf{A_d}, \bm{\alpha})=(y-\mathbf{Z}\bm{\alpha})^T(y-\mathbf{Z}\bm{\alpha})+\lambda \|\bm{\alpha}\|_1,
\end{array}
\end{equation}
in which the data matrix $\mathbf{Z}$ is uniquely defined by the adjacency matrix $\mathbf{A_d}$ and the sampled data $(\mathbf{x}(k), y(k))_{k=1}^{N_s}$.

As it's extremely time-consuming to solve a mixed-integer nonlinear programming problem, here in this paper, the variables $\mathbf{A_d}$ and $\bm{\alpha}$ are optimized by 2 subproblems in turn, i.e., the structure optimization concerning $\mathbf{A_d}$ with $\bm \alpha$ fixed, and the weight optimization problem concerning $\bm \alpha$ with $\mathbf{A_d}$ fixed.
 
\subsubsection{Structure optimization}
For the structure optimization, as the source nodes are provided according to (\ref{source}), hence the first $N_d$ columns of $\mathbf{A_d}$ are fixed to be zero.
 
 For the next $(M-N_d)$ columns, considering the column $j (N_d<j\leq M)$, according to (\ref{generation_strategy}), only $j-1$ elements can be nonzero. Hence the total number of binary variables to be optimized is 
\[
N_d+(N_d+1)+\ldots+M-1=\frac{(N_d+M-1)(M-N_d)}{2},
\]
which is quadratic in $M$. When $M$ is large, it is prohibitive to obtain an optimal structure through the binary programming problem. Hence, a natural thought is to optimize one part of the binary variables at a time and fix the other parts. Then different parts of the binary variables are optimized in turn.  

Specifically, for the column $j$, randomly letting 2 of the $j-1$ elements to be 1, then there will be $j(j-1)/2$ combinations. Traversing all the combinations and choose the one that yields the least cost in (\ref{EHH-optimization}). It is noted that when optimizing the column $j$, the other columns in $\mathbf{A_d}$ are fixed.

\subsubsection{Weights optimization}
For the optimization of the weights $\bm{\alpha}$, the Lasso optimization problem (\ref{lasso-prune}) is solved with respect to the newly obtained adjacency matrix $\mathbf{A_d}$ in the structure optimization. Notice that after weights optimization, $\mathbf{A_d}$ may have changed due to the sparsity of $\bm{\alpha}$.

Iteratively, the optimization of the EHH neural network starts again to optimize the columns $N_d+1, \ldots, M$ in $\mathbf{A_d}$ and then finding the optimal weights $\bm{\alpha}$ given a fixed $\mathbf{A_d}$. The optimization cycles 
can be repeated until the cost function in (\ref{EHH-optimization}) stops deceasing. Algorithm \ref{alg_train} lists the overall optimization procedure. In this paper, we refer to \textbf{1-cycle optimization} of the EHH neural network as a sequent structure optimization and weights optimization, corresponding to line \ref{line-for1} to \ref{line-lasso}.

 \algrrule[0.8pt]
 \begin{alg}
 {Training the EHH neural network.}
 \label{alg_train}
 \algrule[0.5pt]
 \begin{algorithmic}[1]
 \hspace{-4ex} \textbf{Input:} Training data $(\mathbf{x}(k), y(k))_{k=1}^{N_s}$, the preset number of neurons.\\
 \hspace{-4ex} \textbf{Output:} The EHH neural network.
 \STATE Get the q-quantiles in each dimension, say $\beta_{11}, \ldots, \beta_{1,q-1}, \ldots, \beta_{n1}, \ldots, \beta_{n,q-1}$;
 \STATE The neurons in the first hidden layer are listed as (\ref{source}).
  \REPEAT
 \FOR{$j =nq+1:M$ }\label{line-for1}
 \REPEAT
 \STATE Choose 2 elements of $\mathbf{A_d}(i,j)$ with $i<j$ to be 1.\label{line-random}
 \STATE Fixing $\bm{\alpha}$ and calculate the cost in (\ref{EHH-optimization}).\label{line-cal}
 \UNTIL{All the $\frac{j(j-1)}{2}$ combinations are traversed.}
 \STATE Choose the combination that yields the least cost in (\ref{EHH-optimization}).
  \ENDFOR  \label{endfor1}
 \STATE Solve the lasso optimization problem (\ref{lasso-prune}) to get the optimal weights $\bm \alpha$. \label{line-lasso}
  \UNTIL{The cost in (\ref{EHH-optimization}) stops decreasing.}\label{terminate-condition}
 \algrule[0.5pt]
 \end{algorithmic}
 \end{alg}
 
  \subsection{Complexity analysis}
 Before the complexity analysis, we show that  Algorithm \ref{alg_train} is a non-increasing algorithm and can reach a local optimum in neighbours. The definition of neighbours for this particular problem is defined as follows.

 \begin{defn}\label{defn_neighbour}
For the optimization problem (\ref{EHH-optimization}), given the adjacency matrix $\mathbf{A_d}_0$ and weights $\bm{\alpha}_0$, the neighbour  is defined as the set $\mathcal{N}(\mathbf{A_d}_0, \bm \alpha_0)$ such that $\forall (\mathbf{A_d}, \bm{\alpha}) \in \mathcal{N}(\mathbf{A_d}_0, \bm \alpha_0)$, the following holds
\begin{description}
\item[1)] $\mathbf{A_d}$ and $\mathbf{A_d}_0$ only differ in 1 column.
\item[2)] $\|\bm{\alpha}-\bm{\alpha}_0\|\leq \delta$, in which $\delta$ is a scalar small enough.
\end{description}
  \end{defn}
  
   \begin{lem}
 Algorithm \ref{alg_train} is a non-increasing algorithm, i.e.,
  \[
 J(\mathbf{A_d}_1, \bm{\alpha}_1) \leq J(\mathbf{A_d}_0, \bm{\alpha}_0),
 \]
 where $\mathbf{A_d}_0, \bm{\alpha}_0$ and $\mathbf{A_d}_1, \bm{\alpha}_1$ are the adjacency matrices as well as the weights before and after 1-cycle optimization.
 \end{lem}
 \begin{pf}

 For the optimization of the structure, when optimizing the column $\mathbf{A_d}(:, j) (j \in \{N_d+1, \ldots, M\})$, the other columns $\mathbf{A_d}(:, k), k \in \{1, \ldots, M\}, k \neq j$ remain unchanged. Considering the cost 
 \begin{equation}\label{cost_structure}
 J(\mathbf{A_d})=\left(y-\mathbf{Z}(\mathbf{A_d})\right)^T\left(y-\mathbf{Z}(\mathbf{A_d})\right)+\lambda \|\bm{\alpha}_0\|_1,
 \end{equation}
 in which $\bm{\alpha}_0$ is fixed. The column $\mathbf{A_d}_1(:, j)$ is chosen as the one yielding the least cost (\ref{cost_structure}). During this process, only 1 column is optimized at a time. Therefore, after optimizations for the column $\mathbf{A_d}(:, N_d+1), \ldots, \mathbf{A_d}(:, M)$, the cost (\ref{cost_structure}) keeps non-increasing and
 \[
 J(\mathbf{A_d}_1, \bm{\alpha}_0) \leq J(\mathbf{A_d}, \bm{\alpha}_0), \forall (\mathbf{A_d}, \bm \alpha_0) \in \mathcal{N}(\mathbf{A_d}_0, \bm \alpha_0).
 \]

As $\bm{\alpha}_0$ stays constant in the optimization of the structure, it may not be optimal for the optimization problem (\ref{lasso-prune}). Hence, after the structure has been obtained, solving the problem (\ref{lasso-prune}) would further reduce the cost $J(\bm{\alpha})$ in (\ref{lasso-prune}), i.e.,
\[
J(\mathbf{A_d}_1, \bm{\alpha}_1) \leq J(\mathbf{A_d}_1, \bm{\alpha}_0)
\]
 
 Therefore, repeated optimization of the adjacency matrix $\mathbf{A_d}$ and the weights $\bm{\alpha}$ will result in a non-increasing cost $J(\mathbf{A_d}, \bm{\alpha})$ and the algorithm stops at a point with minimum cost in neighbours.
 \end{pf}
 
 The worst-case complexity of 1-cycle optimization, i.e., line \ref{line-for1}-\ref{line-lasso} in Algorithm \ref{alg_train} is given in the following lemma.
 \begin{lem}
 The worst-case complexity for 1-cycle optimization in Algorithm  \ref{alg_train}  is $O(M^4N_s)$.
 \end{lem}
\begin{pf}
In the structure optimization, for the optimization of the column $j$ with $j \in \{N_d+1, \ldots, M\}$, we have to traverse 
\[
\frac{j(j-1)}{2}
\]
combinations in order to find the optimal value of the column during this cycle, corresponding to line \ref{line-random}-\ref{line-cal}.

For each combination, in order to obtain the data matrix $\mathbf{Z}$, we have to conduct $2\cdot N_s$ comparisons for the neuron $j$, then for neurons such that $\mathbf{A_d}_{j k}=1$ with $j <k\leq M$, the corresponding values in the data matrix $\mathbf{Z}$ also have to be changed. Thus we have at most $(M-j)\cdot N_s$ comparisons. After obtaining the data matrix $\mathbf{Z}$, $N_s \cdot (M+1)$ multiplications and $N_s \cdot (M+1)$ additions have to be done to get $\mathbf{Z} \cdot \bm{\alpha}$. Then $N_s$ multiplications and $N_s$ additions are required to get the quantity $(\mathbf{y}-\mathbf{Z}\bm{\alpha})^T(\mathbf{y}-\mathbf{Z}\bm{\alpha})$. Hence in total, for the column $j$, at most
\[
\frac{j(j-1)}{2}\cdot (3M+6)\cdot N_s
\]
mathematical operations are required. Hence the worst-case complexity for the optimization of each column is $O(M^3N_s)$. Considering the optimization for all columns, the worst-case complexity for line \ref{line-for1}-\ref{line-lasso} is 
\[
O(M^4 N_s).
\]

After the structure optimization, the Lasso optimization problem is solved using the ADMM algorithm. The worst-case complexity of ADMM algorithm is $O(1/\epsilon^2)$, where $\epsilon$ is the primal and dual residuals in the ADMM algorithm. Usually a few tens of
iterations is enough to obtain a moderate accuracy, which is sufficient for many large-scale applications \cite{Boyd2011distributed}. Hence in 1-cycle optimization of Algorithm \ref{alg_train}, the complexity of solving a Lasso optimization problem is negligible compared with $O(M^4 N_s)$.

Therefore the overall worst-case complexity for each cycle of optimization is $O(M^4 N_s)$.

\end{pf}

It is clear that the worst-case complexity is high for large $M$. To reduce complexity, we can redefine a smaller neighbour and design corresponding algorithms. For example, instead of optimizing a column at a time, we can only optimize an element  in the column at a time, i.e., pick up $\mathbf{A_d}_{k_1, j}$ and $\mathbf{A_d}_{k_2, j}$ such that 
\[
\mathbf{A_d}_{k_1, j}=1, \mathbf{A_d}_{k_2, j}=1.
\]
And we optimize $k_1$ while fixing $k_2$ and the other columns in $\mathbf{A_d}$ as well as the weights $\bm{\alpha}$. After $k_1$ has been optimized, we optimize $k_2$ while fixing the newly generated $k_1$ and the other columns in $\mathbf{A_d}$ as well as $\bm{\alpha}$. In this case, the mathematical operations for each column $j$ will be reduced to
\[
2(j-1)\cdot (3M+6) \cdot N_s.
\]
Hence the worst-case complexity for each cycle of optimization is $O(M^3N_s)$.

In this situation, the neighbour is smaller by adding the following condition to the conditions 1) and 2),
\begin{description}
\item[3)] $\mathbf{A_d}$ and $\mathbf{A_d}_0$ only differs in 1 element.
\end{description}
 
The number of cycles of optimizations is problem-dependent,  hence the worst-case complexity for Algorithm \ref{alg_train} can not be estimated. In the simulations, generally the number of cycles is around 10, and we will report this in detail in the simulations. For problems with hundreds of dimensions, usually there should be thousands of neurons in order to get a high accuracy, and in this case even the reduced complexity $O(M^3N_s)$ is also high. Currently  we are mainly dealing with problems that only need hundreds of neurons, and extremely large-scale problems are left for our future work.

\section{Nonlinear dynamic system identification using the EHH neural network}
Given a nonlinear dynamic system
\begin{equation}
y(k)=f(\varphi(k), \theta)+\varepsilon(k),
\end{equation}
where ${y}(k)$ denotes the output at time $k$, $f$ is an unknown nonlinear relationship, $\varphi(k)$ is the regression vector and $\varepsilon(k)$ is the error term. The identification goal is to find a nonlinear mapping  such that the additive error term $\varepsilon(k)$ is small.

The nonlinear dynamic system identification problem can be described as,
\begin{equation}
\hat{y}(k)=\hat{f}(\varphi(k)),
\end{equation}
where $\hat{y}(k)$ denotes the predicted output at time $k$, $\hat{f}$ is the approximated nonlinear relationship to be determined. Here we consider the black-box NARX model \cite{Sjoberg1995,Suykens2012artificial}, i.e., the regression vector $\varphi(k)$ consists of the past outputs $y(k-1), \ldots, y(k-n_b)$ and the current as well as past inputs $u(k), u(k-1), \ldots, u(k-n_a)$, 
\begin{equation}\label{regressor}
\varphi(k)=[y(k-1), \ldots, y(k-n_b), u(k), u(k-1), \ldots, u(k-n_a)]^T.
\end{equation}  
Given the input $u(k)$ and the observed output $y(k)$ for $t=1, 2, \ldots, N$, our objective here is to find an optimal EHH neural network as the approximated nonlinear relationship $\hat{f}$. To test the model on the test set, the simulation error is considered \cite{Belz2017automatic}, in which only the input is used to generate the simulated output, that is
\begin{equation}
y_s(k)=\hat{f}(\varphi_s(k)),
\end{equation}
in which 
\begin{equation*}
\varphi_s(k)=[y_s(k-1), \ldots, y_s(k-n_b), u(k), u(k-1), \ldots, u(k-n_a)]^T.
\end{equation*}

  The subsequent benchmark examples will show the effectiveness of the EHH neural networks for the modeling of nonlinear systems, besides, the importance of the regressors as well as the interactions among the regressors will be revealed.

\subsection{Narendra-Li benchmark}

This benchmark was firstly introduced by \cite{Narendra1996}, and has become a benchmark for nonlinear system identification, e.g., \cite{Wen2007,Xu2009ahh}. The benchmark is defined by the state-space form
\begin{equation}\label{narx1-dynamic}
\begin{array}{rl}
x_1(k+1)&=\left(\frac{x_1(k)}{1+x_1^2(k)}+1\right)\sin x_2(k), \\
x_2(k+1)&=x_2(k)\cos x_2(k)+x_1(k)e^{-\frac{x_1^2(k)+x_2^2(k)}{8}} 
+\frac{u^3(k)}{1+u^2(k)+0.5\cos(x_1(k)+x_2(k))},\\
y(k)&=\frac{x_1(k)}{1+0.5\sin x_2(k)}+\frac{x_2(k)}{1+0.5\sin x_1(k)}+e(k).
\end{array}
\end{equation}
According to  \cite{Narendra1996}, this system does not correspond to any real system. It is chosen to be sufficiently complex and nonlinear so that the conventional linear method will not give satisfactory performance. The Gaussian white noise $e(k)$ is generated with a variance of 0.1. The states are assumed not to be measurable. The training signal is obtained with an input sequence sampled uniformly from $[-2, 2]$, with a data length 2000. The test data is generated with the input
\[
u(k)=\sin(2\pi k/10)+\sin(2\pi k/25), k=1, \ldots, 200.
\]

\subsubsection{EHH approximation for the Narendra-Li benchmark}
Following the discussion in \cite{Wen2007,Xu2009ahh}, the regression vector is chosen to be 
\[
\varphi(k)=[y(k-1), y(k-2), y(k-3), u(k-1), u(k-2), u(k-3)]^T.
\]
The same as \cite{Wen2007}, the performance of the approximation is assessed by the test VAF (variance- accounted-for), i.e.,
\[
\mathrm{VAF}(\hat{f})=\max\left\{0, 1-\frac{\mbox{var}(\hat{f}-y)}{\mbox{var}(y)}\right\},
\]
in which $\hat{f}$ denotes the approximations of the true system using different kinds of methods and var represents the variance.


To determine the number of source nodes $N_d$ and the number of total neurons $M$, we test 9 groups of values, i.e., 
\[N_d = 6\times 5, 6 \times 10, 6 \times 20,\]
and 
\[M=N_d+30, N_d+40, N_d+50.\] 
The number of parameters is not large as the dimension is not high. After training 9 different EHH neural networks, we select the one with the least GCV criterion. The corresponding structure is set to be the network structure in subsequent experiments. For this problem, the number of source nodes and intermediate nodes are preset to be $6 \times 5$ and 40, respectively.

10 experiments are then performed and the one resulting the best test RSSE is selected to compare with the other methods.
The estimation procedure includes 5 cycles of evaluating line \ref{line-for1} to line \ref{line-lasso}, and the elapsed time is 2.83s. The number of neurons employed finally is 59, which is smaller compared with the preset 70.  All the computations in this paper are implemented through MATLAB 2016b on a 2.7 GHz Intel Core i7 computer. 

\begin{figure}[htbp]
\begin{center}
\psfrag{y}[c]{$y$}
\psfrag{t}[c]{$t$}
\subfigure[]{
\includegraphics[width=0.48\columnwidth]{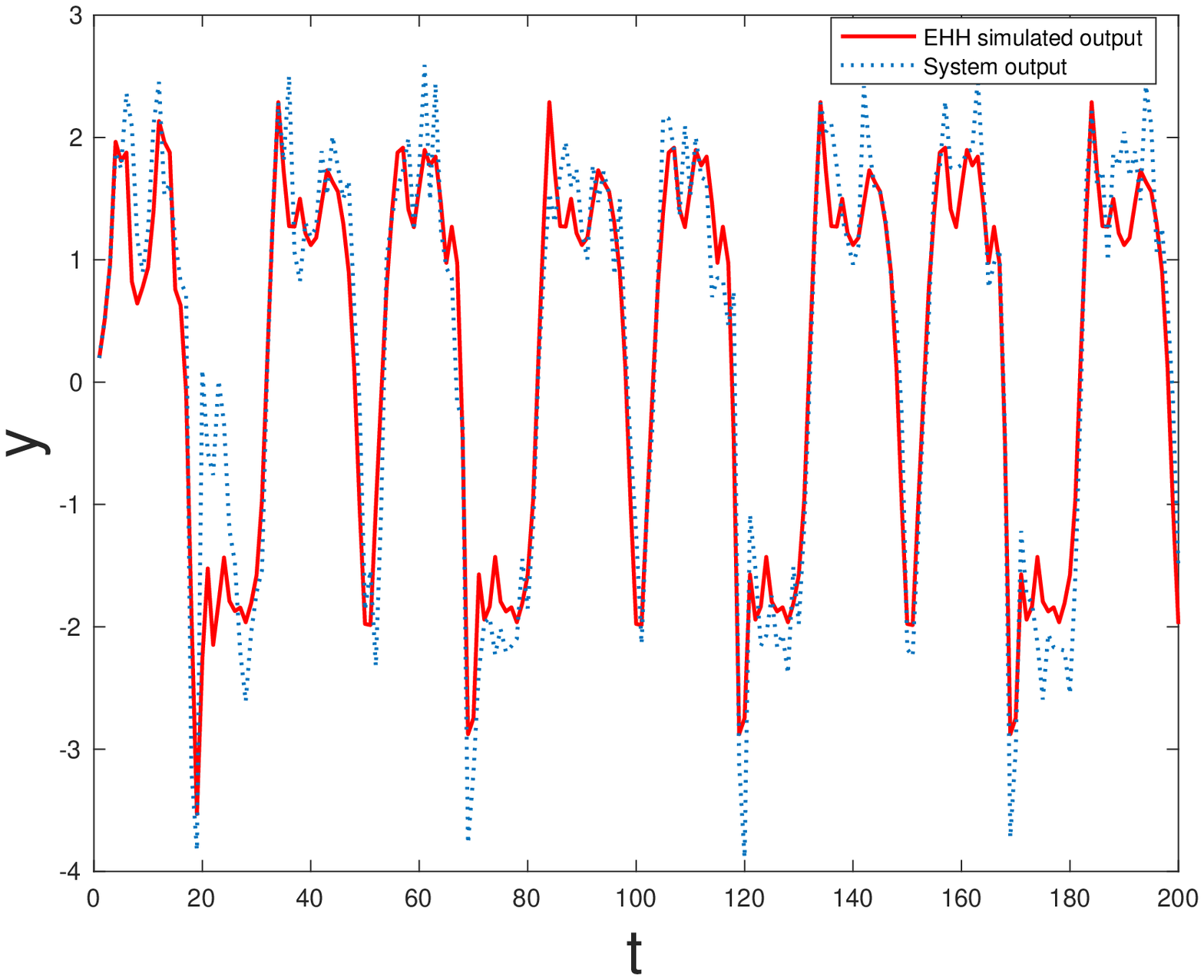}
\label{fig-narx1_ehh}}
\subfigure[]
{\includegraphics[width=0.48\columnwidth]{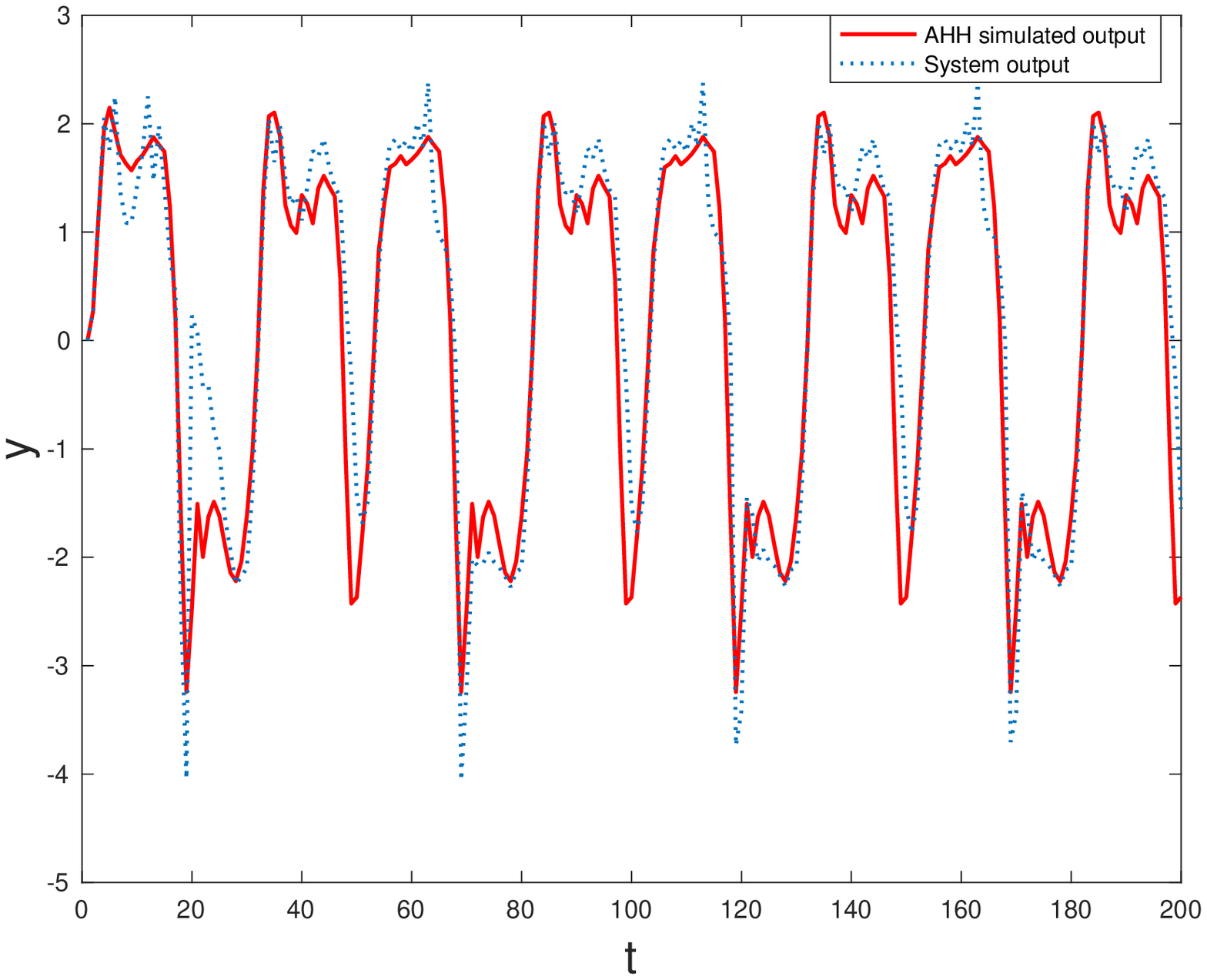}
\label{fig-narx1_ahh}}
\caption{Test performance of the EHH neural network and the AHH tree for the Narendra-Li benchmark. (a) The EHH simulated output and the system output. (b) The AHH simulated output and the system output.}
\label{fig-narx1}
\end{center}
\end{figure}

The simulated output of the EHH network is shown in red solid in Fig. \ref{fig-narx1_ehh}, in which the system output  is also shown (in blue dotted). The simulated output of the AHH model is taken from \cite{Xu2009ahh} and shown in red solid in Fig. \ref{fig-narx1_ahh} for comparison, in which again the system output is shown in blue dotted.

The test VAF (simulated VAF),  the calculation complexity and the number of parameters (denoted as \# Parameters) of the EHH neural network and the corresponding initial neural network are shown in Table \ref{tab-narx1}. The results for the NARX model with AHH model \cite{Xu2009ahh}, basis piecewise affine (BPWA) \cite{Wen2007} are also listed for comparison. The computational complexity for the AHH model is re-evaluated through the current computer while that for the BPWA strategy is not available, and shown as ``-".
 \begin{table}[htbp]
\begin{center}
\caption{Comparison of test performance of AHH model, the BPWA strategy and EHH neural network for the Narendra-Li system.}
\label{tab-narx1}
\begin{tabular}{ccccc}
\hline
{}&AHH \cite{Xu2009ahh}&BPWA \cite{Wen2007}&EHH$_{\mathrm{ini}}$&EHH\\
\hline
RSSE&87.4\%&89.1\%&0&90.5\%\\
\hline
Elapsed Time&49.8&-&0.64s&2.83s\\
\hline
\# Parameters & 132&847&187&178\\
\hline
\end{tabular}
\end{center}
\end{table}

It can be seen from Table \ref{tab-narx1} that the EHH neural network
 gives a good approximation of the original dynamic system (\ref{narx1-dynamic}) and outperforms the strategy of the AHH model and BPWA strategy. 
 Compared with the AHH model, the elapsed time for the EHH neural network is much shorter while the accuracy is better. It is noted that the number of parameters employed for the EHH neural network is a little larger than that for the AHH tree, indicating the effectiveness of the AHH training procedure, although the training lasts for a longer time.

It is also noted that compared with the initial EHH neural network, the final EHH neural network gives a much better accuracy and at the same time, the number of parameters is reduced, verifying the effectiveness of the structure and weights optimization.

\subsubsection{ANOVA decomposition of the EHH approximation}

The ANOVA functions with the top 5 $\sigma$-values  of the resulting EHH neural network are listed in Table \ref{anova-narx1}. Larger $\sigma$-value indicates that the ANOVA function is more important. The term $\setminus$GCV for an ANOVA function represents the GCV score for a model with all of the nodes corresponding to that particular ANOVA function removed. Hence it can
provide  another indication of the importance of the corresponding ANOVA function, i.e., larger $\setminus$GCV indicates the more importance of the ANOVA function.
\begin{table}[h]
\begin{center}
\caption{The ANOVA functions with the top 5 $\sigma$-values of the EHH neural network for the Narendra-Li system.}
\label{anova-narx1}
\begin{tabular}{cccc}
\hline
ANOVA Fun.& Regressor & $\sigma$& $\setminus$GCV
\\ \hline
1&$y(k-1), y(k-2), u(k-2)$&1.7080&3.8234\\
2& $u(k-1)$ &1.2001&3.6748\\
3&$y(k-1), y(k-2)$&0.8815&1.0973\\
4&$y(k-2)$&0.7668 &2.9347 \\
5&$y(k-1)$&0.5451 &0.6632\\
\hline
\end{tabular}
\end{center}
\end{table}

From the table we can know that the trend of the $\sigma$-value and $\setminus$GCV are approximately the same. The most important contribution comes from the interaction among $y(k-1), y(k-2), u(k-2)$, and the additive terms $y(k-1), y(k-2), u(k-1)$ also contribute largely. 

It is noted that for the NARX based system identification, the ANOVA decomposition only gives a rough indication of the importance of the regressors as well as the regressor interactions, as different regressors are actually dependent on each other. 

\subsection{Bouc-Wen system}
The Bouc-Wen model has been intensively used to model hysteretic effects in mechanical engineering, especially in the case of random vibrations. An extensive literature review about the Bouc-Wen modeling can be found in \cite{Ismail2009hysteresis}. The Bouc-Wen benchmark system has been described in detail in \cite{Noel2016hysteretic,Westwick2018using}. The training signal consists of 5 periods of a multisine and has a total length of 40960 samples.  The first 2 periods of this signal were removed for the transient \cite{Esfahani2018parameter}. There are 2 sets of test data, one is with multisine input and the other is with swept-sine input, with lengths 8192 and 153000, respectively.

\subsubsection{EHH approximation of Bouc-Wen system}
The same as \cite{Westwick2018using}, the regressor is chosen to be
\[
\varphi(k)=[y(k-1), \ldots, y(k-15), u(k), \ldots, u(k-14) ]^T.
\]

And according to \cite{Noel2016hysteretic}, the performance of the approximation is judged by the test RMSE (root mean square error), i.e.,
\[
\mathrm{RMSE}(\hat{f})=\sqrt{\frac{1}{N}\sum\limits_{k=1}^N(\hat{f}(k)-y(k))^2},
\]
in which $\hat{f}$ denotes the approximations of the true system using different kinds of methods.

To determine the number of source nodes $N_d$ and the number of total neurons $M$, as the dimension for this problem is much higher than the Narendra-Li benchmark, we test the following 9 groups of values, i.e., 
\[N_d = 30\times 5, 30 \times 10, 30 \times 20,\]
and 
\[M=N_d+100, N_d+200, N_d+250.\] 
After training 9 different EHH neural networks, we select the one with the least GCV criterion. The corresponding structure is set to be the network structure in subsequent experiments. For this problem, the number of the source nodes and intermediate nodes are preset to be $30 \times 5$ and 200, respectively.

Then 10 experiments are done, and the one with the best test RMSE is shown 
in Table \ref{tab-bouc-wen}, in which the result of the initial EHH neural network (without the structure and weights optimization) is also listed. The time complexity for getting an initial EHH neural network is approximately 8.6s, then 305s is used to get the final EHH neural network. The whole procedure includes 5 cycles of structure and weights optimization, and the final EHH neural network includes 145 neurons.  The error is listed in dB form, i.e., $ 20\log 10(\mathrm{RMSE})$. The results for the NARX model with tree-based local model networks \cite{Belz2017automatic}, Volterra Feedback Model \cite{Schoukens2016modeling}, and decoupled polynomial NARX model \cite{Westwick2018using} are also listed for comparison. It is noted that the application of the AHH model in the Bouc-Wen benchmark  problem is prohibitive due to the high dimension (30 dimensions).


  \begin{table}[htbp]
\begin{center}
\caption{Comparison of test performance of several approaches on Bouc-Wen system.}
\label{tab-bouc-wen}
\begin{tabular}{ccccccc}
\hline
{}&NARX \cite{Belz2017automatic}&Volterra Feedback \cite{Schoukens2016modeling}&Decoupled NARX \cite{Westwick2018using}&EHH$_{\mathrm{ini}}$&EHH\\
\hline
RMSE (multisine)&-75.73&-81.51&-85.42&-84.56&\textbf{-86.11}\\
\hline
RMSE (swept-sine)&-77.20&-85.03&\textbf{-95.55}&-90.09&-92.39 \\
\hline
\# Parameters&-&-&226&610 &436\\
\hline
\end{tabular}
\end{center}
\end{table}

It can be seen from Table \ref{tab-bouc-wen} that the EHH neural network performs well, and outperforms the procedure in \cite{Belz2017automatic} and \cite{Schoukens2016modeling}. We can also see that the performance of the optimized EHH neural network is better than the initial EHH neural network while at the same time, the number of parameters is reduced, verifying the effectiveness of the structure and weights optimization. 
Compared with the decoupled NARX \cite{Westwick2018using}, the RMSE for the multisine test data is smaller while that for the sinesweep data is a little larger. 

Besides, the number of parameters employed in the EHH neural network  is 436, while in \cite{Westwick2018using}, the number of parameters is 226. 
 To further improve the accuracy and reduce the number of parameters of the EHH neural network, we can also employ a linear transformation before the input layer like that in \cite{Westwick2018using}, which somewhat reflects the physical insights of the dynamic system. And this part of work has been finished as our another work \cite{Liang2019decoupled}. Or we can build a state-space version for the EHH neural network, like \cite{Suykens1995nonlinear}, which we will consider in the future.

\begin{figure}[htbp]
  \centering
  \subfigure[]{
    \label{bouc1} 
    \includegraphics[width=0.45\columnwidth]{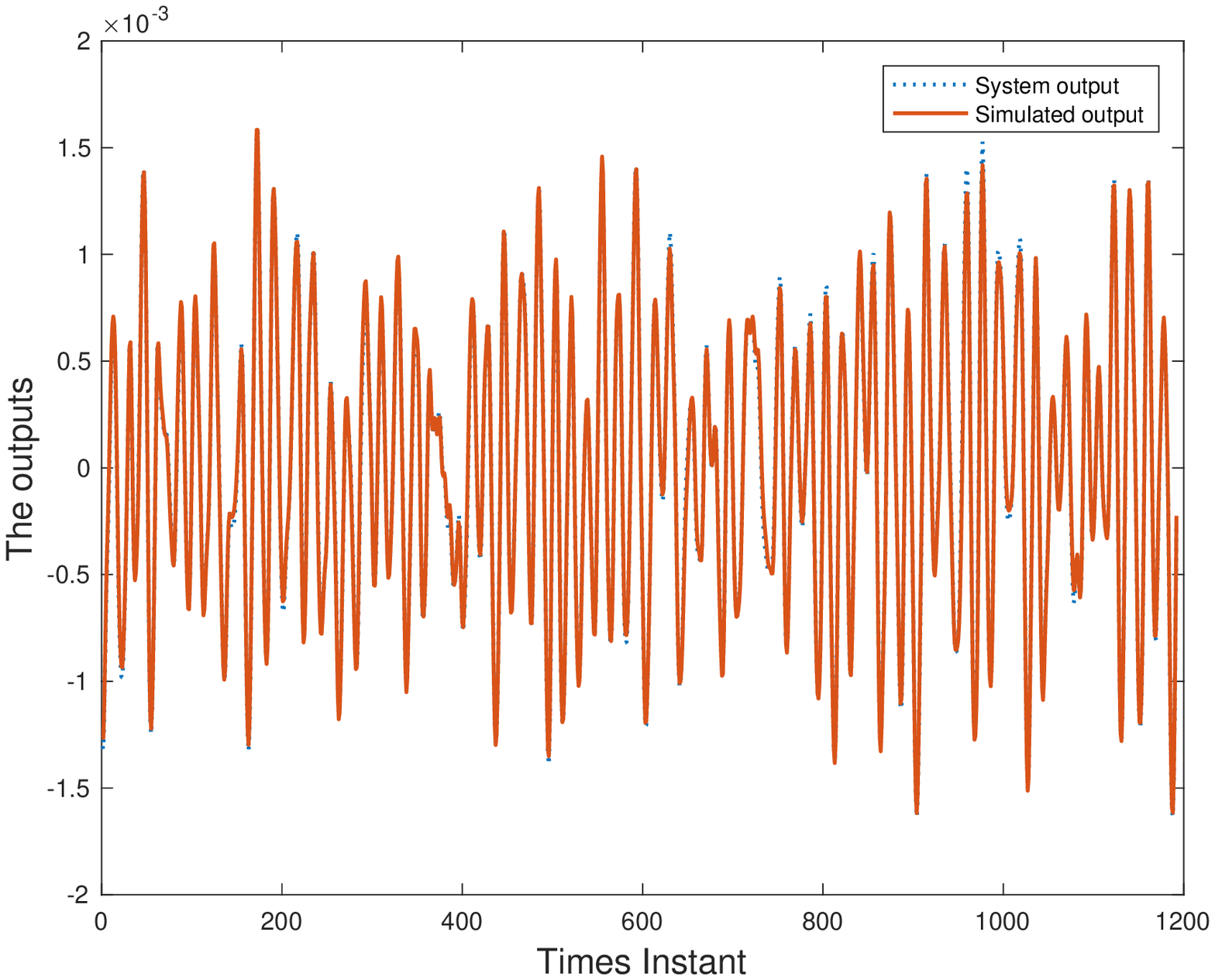}}
    \subfigure[]{
    \label{bouc2} 
    \includegraphics[width=0.45\columnwidth]{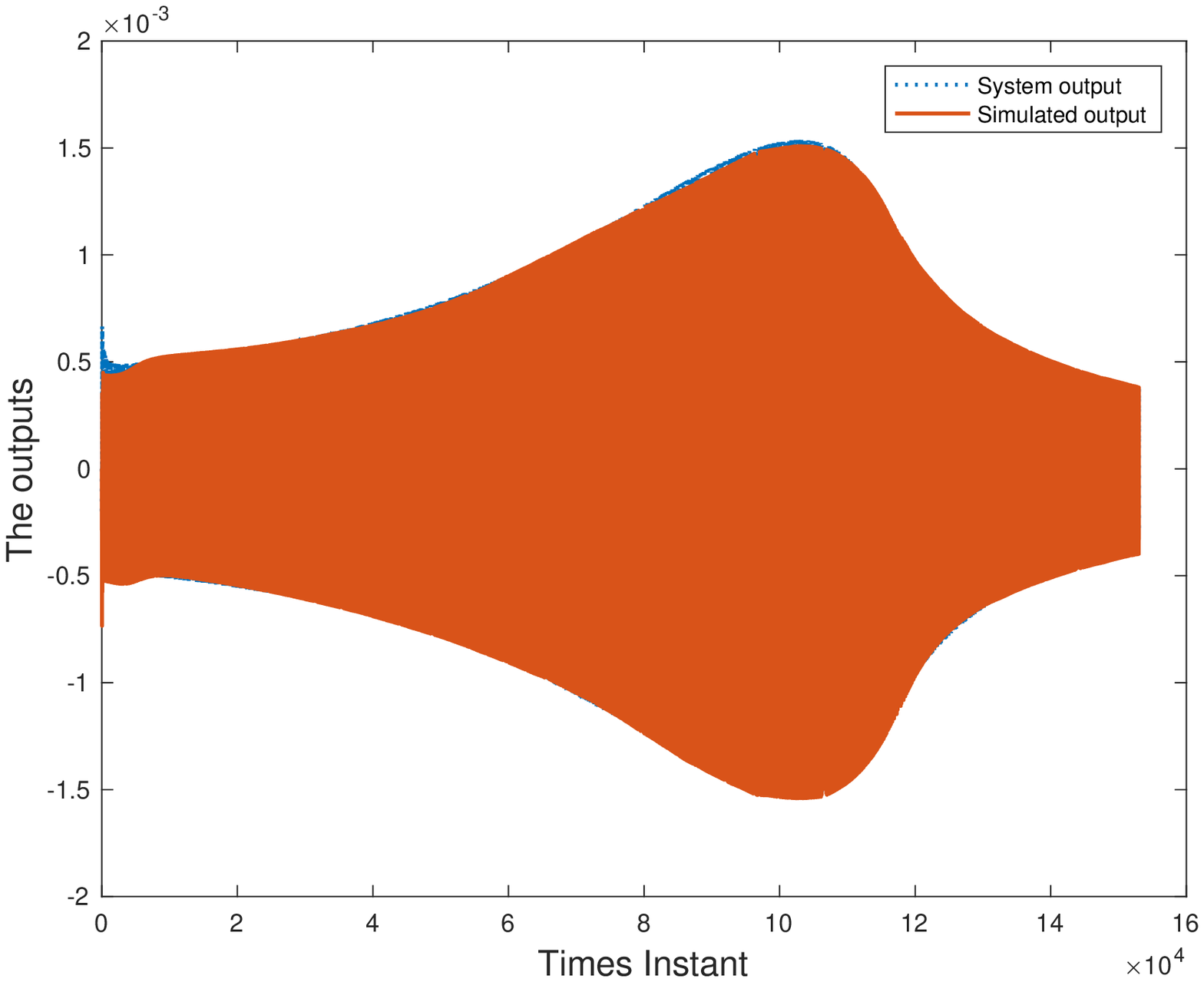}}
         \caption{The simulated performance of the EHH neural network for the Bouc-Wen benchmark. (a) The simulated outputs of the EHH neural network and the system output with multisine input. (b) The simulated outputs of the EHH neural network and the system output with swept-sine input.}
          \label{bouc-wen-figure}
      \end{figure}

Fig. \ref{bouc-wen-figure} shows the simulated output (shown in red solid) as well as the system output (shown in blue dotted), in which Fig. \ref{bouc1} only selects the last 1200 points to provide a clear illustration.  It can be seen from Fig. \ref{bouc-wen-figure} that the simulated output approaches the system output quite well.

  \subsubsection{ANOVA decomposition of the EHH approximation}
 From the EHH NARX model, the contributions of regressors as well as the interactions among different regressors are clear. Here as the input dimension is high, the ANOVA functions with the top 10 $\sigma$-values of the resulting EHH neural network is listed in Table \ref{anova-narx2}.
 \begin{table}[h]
\begin{center}
\caption{The ANOVA functions with the top 10 $\sigma$-values of the EHH neural network for the Bouc-Wen example.}
\label{anova-narx2}
\begin{tabular}{cccc|cccc}
\hline
ANOVA Fun.& Regressor & $\sigma $& $\setminus \mathrm{GCV}$&ANOVA Fun.& Regressor & $\sigma $& $\setminus \mathrm{GCV}$\\
 \hline
1&$y(k-1)$&$8.13 \times 10^{-4}$&15.16&6&$u(k-7)$&$2.69 \times 10^{-5}$&$2.07 \times 10^{-2}$\\
2& $y(k-4)$ &$1.38 \times 10^{-4}$&0.45&7&$y(k-11)$&$2.40 \times 10^{-5}$&$1.35 \times 10^{-2}$\\
3&$y(k-5)$&$6.06 \times 10^{-5}$&$8.64 \times 10^{-2}$&8&$y(k-13)$&$2.10 \times 10^{-5}$&$1.19 \times 10^{-2}$\\
4&$u(k-2)$&$5.68 \times 10^{-5}$&$9.19\times 10^{-2}$&9&$y(k-10)$&$1.74 \times 10^{-5}$&$7.88 \times 10^{-3}$\\
5&$u(k-1)$&$2.86 \times 10^{-5}$&$2.23 \times 10^{-2}$&10&$y(k-1), y(k-9)$&$1.31 \times 10^{-5}$&$1.30 \times 10^{-3}$\\
\hline
\end{tabular}
\end{center}
\end{table}

From Table \ref{anova-narx2}, we can know that the most recent output $y(k-1)$ contributes to the current output most, besides, there is input delay in this system, as the most important input is $u(k-2)$ and $u(k-1)$. From the regressors contained in the ANOVA functions, we can know that the regressors influence the output mostly additively.

\section{Conclusions}
\subsection{Conclusions}
In this work, the efficient hinging hyperplanes (EHH) neural network is proposed, which can be used for function approximation and dynamic system identification. The EHH neural network belongs to the HH family, and for every EHH neural network, an equivalent AHH tree exists. Besides, the EHH neural network is a distributed representation, hence can be used for large-scale problems. After an initial EHH neural network has been obtained, the EHH neural network can be trained through the structure and weights optimization, in which convex optimization problems are solved sequentially. The interpretability of the EHH neural network can be clearly obtained through the interaction matrix and the ANOVA decomposition. The EHH neural network is applied in dynamic system identification, which can be used as a  tool for system approximation and an indication of regressor importance. Simulation results shows the effectiveness of the proposed procedure.

\subsection{Future work}
Although in theory, the EHH neural network is a distributed representation and can be applied to large-scale problem,
the application to problems with hundreds of dimensions is limited due to the efficiency of the present algorithm. Therefore, we are dedicated to improving the current EHH neural network on solving extremely large-scale problems and at the same time, reduce the number of parameters employed.

\begin{ack}
This work is jointly supported by the National Natural Science Foundation of China (U1813224), Chinese Scholarship Council (CSC) and Science and Technology Innovation Committee of Shenzhen Municipality (JCYJ2017-0811-155131785).
Johan Suykens acknowledges support of ERC Advanced Grant E-DUALITY
(787960), KU Leuven C1, FWO G0A4917N.
\end{ack}


\end{document}